\documentclass[a4paper,10pt]{article}
\usepackage[utf8]{inputenc}
\usepackage{mathtools}
\usepackage{amssymb}
\usepackage{mathrsfs}
\usepackage{hyperref}
\usepackage{graphicx}
\usepackage{verbatim}
\usepackage{geometry}
\usepackage{relsize}
\usepackage{array}
\usepackage{tikz}
\usetikzlibrary{patterns}
\tikzstyle{every picture}=[level distance = 8mm, baseline=-0.5ex]
\tikzstyle{prop}=[shape=circle,minimum size=6mm, draw=black!80, fill=green!30]

\newcommand{\SDe}{Schwinger--Dyson equations}
\newcommand{\RGe}{renormalization group equations}
\newcommand{\D}{\mathcal{D}}

\begin{document}

\setlength{\extrarowheight}{3pt}

\title{Ward--Schwinger--Dyson equations\\ in \(\phi^3_6\) Quantum Field Theory}
\author{ Marc~P.~Bellon, Enrico~I.~Russo\\ 
\normalsize \it Sorbonne Université, CNRS, Laboratoire de Physique Théorique et Hautes
Energies,\\ \normalsize \it LPTHE, 75005 Paris, France}

 \date{}    

\maketitle

\begin{abstract}
We develop a system of equations for the propagators and three-point functions of the
\(\phi^3\) quantum field theory in six dimensions.  Inspired from a refinement by Ward on
the Schwinger--Dyson equations, the main characteristics of this system are to be
formulated purely in terms of renormalized quantities and to give solutions satisfying
renormalisation group equations.  These properties were difficult to get together, due to
the overlapping divergences in the propagator. The renormalisation group equations are an integral part
of any efficient resolution scheme of this system and will be instrumental in the study of
the resurgent properties of the solutions.

It is our belief that this method can be generalized to the case of gauge fields, shedding
some light on their quantum properties. 

\end{abstract}

\textbf{Mathematics Subjects Classification:} 81T10, 81T15, 81Q40.


\section*{Introduction}\label{introduction}
In this work, we introduce a system of generalized \SDe\ for a massless $\phi^3$ model in
$6$ dimensions.

Usual works on \SDe\ miss one of their fundamental interest for us, the possibility to solve them purely in terms of
renormalized Green functions.  This possibility was first encountered in the case of a linear
Schwinger--Dyson equation in~\cite{BrKr99}, developed in~\cite{KrYe2006} and found a powerful illustration
in~\cite{BeSc08}, from which a whole series of refinements have emerged, see e.g.,
\cite{Be10a,BeSc12,BeCl13}.

The \SDe\ start from the implication of field equations on the field correlators and involve at first a bare
vertex, but in many cases can be converted to ones depending only on dressed vertices. 
However, this is not possible for the propagator corrections, which are written
asymmetrically, with a bare vertex on one side and a dressed vertex on the other side.
In the language of renormalisation Hopf algebras, the resulting combinatorial Schwinger--Dyson
equation is not based on a Hopf algebra cocycle. It has been proposed to put suitable combinatorial factors
to ensure this cocycle property~\cite{Kr2005}, but this does not solve all problems. First of all,
as has been remarked by K.~Yeats in her thesis~\cite{Ye08}, this cannot solve the problem in
theories where different particles can run in the one loop corrections for a single propagator, as
in QCD. Furthermore, even in cases where such a formulation is valid at the combinatorial level, the
dependence of the combinatorial factors on the internal structure of the graphs precludes its transformation
into an analytic Schwinger--Dyson equation, where one simply uses the values up to some order of the vertex
functions in the right-hand side of the equation.
The problem is directly linked to the presence of overlapping divergences.  The \SDe\ build
complex diagrams by assembling simpler parts, but in the presence of overlapping divergences, the
obtained diagram has alternative ways of being dismantled.  Since each of the ways to
disassemble a diagram implies a possible chain of counterterms, the
counterterms associated to the constituents are not sufficient for a full preparation of the diagram:
subdivergences remain.

The solution we adopt has quite a
long history, since we found its first expression in a paper by J.C.~Ward~\cite{Wa51}, completed
in~\cite{Ward_1951}.  The most complete exposition in these early times appears in a conference report of
Symanzik~\cite{Sy61} and some further applications of these ideas appear for example in~\cite{MiYa66} and
especially~\cite{BaLe77}, where the case of QCD is worked out.  However the complexity of the proposed
solution did not really allow for applications.  An important further advantage of these methods is that they
are naturally compatible with the Ward or Slavnov--Taylor identities expressing gauge invariance, when no
consistent truncation of the usual \SDe\ can be found with this property.  This will not appear in this
paper, where we limit ourselves to scalar interactions, but we hope to go back to this question in future
works.

We have two additional ingredients in our proposal.  First, since the general analytic dependence of
three-point functions on the kinematic invariants rapidly becomes cumbersome, we will reduce to the use of a
single-scale version, which has the same type of functional dependence as the propagator. We will
sketch a systematic computation of the corrections to this first approximation.  The second one is that
\RGe\ are a direct consequence of our Ward--\SDe. The proof we give also applies to the Wess--Zumino model
considered in previous works and is much simpler than the one used in~\cite{BeSc08}: it does not use the Hopf
subalgebra property of the corresponding combinatorial \SDe\ and is independent on previous works on renormalisability.
Since the \RGe\ were an essential part in our analysis of the asymptotic properties of the perturbative
solutions in the case of the Wess--Zumino model, we surmise that the same kind of analysis can be done
for the singularities of the Borel transform as in~\cite{BeCl14} and that, as in~\cite{BeSc12}, higher-order
terms in the Ward--\SDe\ only bring higher-order corrections to the properties of these singularities.

The rest of the paper is structured as follows: in section §\ref{Schwinger-Dyson equations} we will present
the \SDe\ we will be studying and introduce the problem of overlapping divergences that until now has
precluded the use of \SDe\ in the presence of vertex corrections; in section §\ref{IR_rearrangement}
following~\cite{BrKr2013,BrKr2012} we will present a deformation of Feynman rules that resembles the IR
rearrangement used in QCD to have single-scale vertices, paving the way to treat them as solutions of
renormalisation equation; section §\ref{Schwinger-Dyson and renormalization group equations} is a curious
intermezzo that tangles \SDe\ and \RGe\ providing us with a recipe for the $\beta$-function; and in section
§\ref{First primitive diagram approximation} and §\ref{IRcorrections} we will provide results for the
various anomalous dimensions and thus reconstructing up to order $a^2$ the $2$-point and $3$-point functions; we then conclude.

\section{The Ward--Schwinger--Dyson equations}\label{Schwinger-Dyson equations}

After the introduction of Hopf algebra methods to deal with the combinatorics of renormalisation, it has been
recognized that \SDe\ can be formulated in terms of 1-cocycles \(B_+\) in Hochschild
cohomology~\cite{BrKr2000,Kr2003}.
In the case of the Hopf algebra of (decorated) trees, \(B_+(F)\) is the
tree made by putting the roots of the trees in the forest \(F\) as direct descendants of a new root
(which in the decorated case, will have the decoration associated to \(B_+\)).  One obvious question is
whether the solution of the said \SDe\ satisfies \RGe.  An important stepping stone is to know whether the
series coefficients of the solution would generate a sub Hopf algebra of the Hopf algebra of graphs.  The
possible forms of such \SDe\ were investigated in~\cite{Foi2014} and one of the possible class of \SDe\ seems to
correspond to the situation in quantum field theory. They indeed have variables which can be interpreted as propagators or
vertex functions and the different \(B_+\) act on products of powers of the variables, with the
exponents corresponding to the number of vertices and propagators in a diagram. 

However, Feynman rules for the evaluation do not apply to rooted trees but to Feynman diagrams which do not
always have a nice translation in a tree structure, due to the overlapping divergences.
To make things more concrete, 
we start from the \SDe\ for our model $\phi^3_6$ : 
\begin{eqnarray}\label{propagator}
\begin{tikzpicture}
\draw (-0.5,0) -- (0.5,0);
\filldraw [gray] (0,0) circle (3pt);
\end{tikzpicture}
&=& 
\begin{tikzpicture}
\draw (-0.5,0) -- (0.5,0);
\end{tikzpicture}
- \frac{1}{2}\;\;
\begin{tikzpicture}
\draw (-0.8,0) -- (-0.5,0);
\draw (0.5,0) -- (0.7,0);
\draw (0.5,0) arc (0:-180:0.5);
\draw (0.5,0) arc (0:180:0.5);
\filldraw [gray] (-0.5,0) circle (3pt);
\filldraw [black] (0, 0.5) circle (2pt);
\filldraw [black] (0,-0.5) circle (2pt);
\end{tikzpicture}
\\
\label{vertex}
\begin{tikzpicture}
\draw (-0.5,0) -- (0,0);
\draw (0,0) -- (0.5,0.5);
\draw (0,0) -- (0.5,-0.5);
\filldraw [gray] (0,0) circle (3pt);
\end{tikzpicture}
&=&
\begin{tikzpicture}
\draw (-0.5,0) -- (0,0);
\draw (0,0) -- (0.5,0.5);
\draw (0,0) -- (0.5,-0.5);
\end{tikzpicture}
+\;\;
\begin{tikzpicture}
\draw (-0.8,0) -- (-0.5,0) -- (0,0.5) -- (0.8, 0.5);
\draw (-0.5,0) -- (0,-0.5)-- (0.8, -0.5);
\filldraw [gray] (-0.5,0) circle (3pt);
\filldraw [gray] (0.4,0) ellipse (0.2 and 0.6);
\filldraw [black] (0,0.5) circle (2pt);
\filldraw [black] (0,-0.5) circle (2pt);
\end{tikzpicture}
\end{eqnarray}
where the black and gray elements denote, respectively, dressed  propagators and vertices. 
The big oval denotes a four-particle kernel, the Bethe--Salpeter kernel, which can be expanded as
a sum of two-particle irreducible graphs. In the following equation, the internal lines denote the full
propagator, without an additional dot for the sake of readability:
\begin{equation}\label{4ptfunction}
\begin{tikzpicture}
\draw (-0.4,0.5) -- (0.4,0.5) ;
\draw (-0.4,-0.5) -- (0.4,-0.5) ;
\fill[gray] (0,0) ellipse (0.2 and 0.6);
\end{tikzpicture}
=
\begin{tikzpicture}
\draw (-0.4,0.5) -- (0.4,0.5) ;
\draw (-0.4,-0.5) -- (0.4,-0.5) ;
\draw (0,0.5) -- (0,-0.5);
\filldraw [gray] (0,0.5) circle (3pt);
\filldraw [gray] (0,-0.5) circle (3pt);
\end{tikzpicture}
+
\frac{1}{2}
\begin{tikzpicture}
\draw (-0.5,0.5) -- (0.5,0.5) ;
\draw (-0.5,-0.5) -- (0.5,-0.5) ;
\draw (0.2,0.5) -- (-0.2,-0.5);
\filldraw [white] (0,0) circle (3pt);
\draw (-0.2,0.5) -- (0.2,-0.5);
\filldraw [gray] (-0.2,0.5)circle (3pt);
\filldraw [gray] (0.2,0.5)circle (3pt);
\filldraw [gray] (0.2,-0.5)circle (3pt);
\filldraw [gray] (-0.2,-0.5)circle (3pt);
\end{tikzpicture}
+
\begin{tikzpicture}
\draw (-0.5,0.5) -- (0.5,0.5) ;
\draw (-0.5,-0.5) -- (0.5,-0.5) ;
\draw (0.2,0.5) -- (0.2,-0.5);
\draw (-0.2,-0.5) -- (-0.2,0.5);
\draw (-0.2,0) -- (0.2,0);
\filldraw [gray] (-0.2,0.5)circle (3pt);
\filldraw [gray] (0.2,0.5)circle (3pt);
\filldraw [gray] (-0.2,0.)circle (3pt);
\filldraw [gray] (0.2,0.)circle (3pt);
\filldraw [gray] (0.2,-0.5)circle (3pt);
\filldraw [gray] (-0.2,-0.5)circle (3pt);
\end{tikzpicture}
+
\cdots
\end{equation}
The factor before the second diagram is a symmetry factor linked to the invariance of this diagram
through the exchange of its two outputs.  Since we will only make explicit computations at rather low order, the first term will be generally
sufficient.  Combining the three previous equations, one may obtain solutions as series of one-particle
irreducible graphs, but it is not the path we will follow.

The asymmetry in the first equation might puzzle at first glance. Indeed, it is not possible to write an
equation with a single diagram and a constant combinatorial factor.
To restore its symmetry, we use the expression in~\eqref{vertex} at the cost of introducing a second object:
\begin{equation}\label{SDeq_compl}
\begin{tikzpicture}
\draw (-0.5,0) -- (0.5,0);
\filldraw [gray] (0,0) circle (3pt);
\end{tikzpicture}
= 
\begin{tikzpicture}
\draw (-0.5,0) -- (0.5,0);
\end{tikzpicture}
- \frac{1}{2}\;\;
\begin{tikzpicture}
\draw (-0.8,0) -- (-0.5,0);
\draw (0.5,0) -- (0.8,0);
\draw (0.5,0) arc (0:-180:0.5);
\draw (0.5,0) arc (0:180:0.5);
\filldraw [gray] (-0.5,0) circle (3pt);
\filldraw [gray] (0.5,0) circle (3pt);
\filldraw [black] (0, 0.5) circle (2pt);
\filldraw [black] (0,-0.5) circle (2pt);
\end{tikzpicture}
+ \frac{1}{2}\;\;
\begin{tikzpicture}
\draw (-0.8,0) -- (-0.5,0) -- (0,0.5) -- (0.8, 0.5) --  (1.3, 0) --(1.6,0);
\draw (-0.5,0) -- (0,-0.5)-- (0.8, -0.5)--  (1.3, 0);
\filldraw [gray] (-0.5,0) circle (3pt);
\filldraw [gray] (1.3,0) circle (3pt);
\fill[gray] (0.4,0) ellipse (0.2 and 0.6);
\filldraw [black] (-0.2,0.3) circle (2pt);
\filldraw [black] (-0.2,-0.3) circle (2pt);
\filldraw [black] (1,0.3) circle (2pt);
\filldraw [black] (1,-0.3) circle (2pt);
\end{tikzpicture}
\quad .
\end{equation}
Each term in this equation produces multiple counting as can be seen already from the first few terms
of their expansion:
\begin{align*}
\hspace{-0.3cm}
\begin{tikzpicture}
\draw (-0.8,0) -- (-0.5,0);
\draw (0.5,0) -- (0.8,0);
\draw (0.5,0) arc (0:-180:0.5);
\draw (0.5,0) arc (0:180:0.5);
\filldraw [gray] (-0.5,0) circle (3pt);
\filldraw [gray] (0.5,0) circle (3pt);
\filldraw [black] (0, 0.5) circle (2pt);
\filldraw [black] (0,-0.5) circle (2pt);
\end{tikzpicture}
\,
=
\,
\begin{tikzpicture}
\draw (-0.6,0) -- (-0.5,0);
\draw (0.5,0) -- (0.6,0);
\draw (0.5,0) arc (0:-180:0.5);
\draw (0.5,0) arc (0:180:0.5);
\end{tikzpicture}
\,
+
\, 
2
\,
 &
\begin{tikzpicture}
\draw (-0.6,0) -- (-0.5,0);
\draw (0.5,0) -- (0.6,0);
\draw (0, 0.5) -- (0, -0.5);
\draw (0.5,0) arc (0:-180:0.5);
\draw (0.5,0) arc (0:180:0.5);
\end{tikzpicture}
\,
+
\,
 3
\,
\begin{tikzpicture}
\draw (-0.6,0) -- (-0.5,0);
\draw (0.5,0) -- (0.6,0);
\draw (0.15, 0.475) -- (0.15, -0.475);
\draw (-0.15, 0.475) -- (-0.15, -0.475);
\draw (0.5,0) arc (0:-180:0.5);
\draw (0.5,0) arc (0:180:0.5);
\end{tikzpicture}
\,
+
\,
2
\,
\begin{tikzpicture}
\draw (-0.6,0) -- (-0.5,0);
\draw (0.5,0) -- (0.6,0);
\draw (-0.35,-0.355) -- (0.35, 0.355);
\filldraw [white] (0,0) circle (3pt);
\draw (-0.35,0.355) -- (0.35, -0.355);
\draw (0.5,0) arc (0:-180:0.5);
\draw (0.5,0) arc (0:180:0.5);
\end{tikzpicture}
+ \cdots,
\\
\begin{tikzpicture}
\draw (-0.8,0) -- (-0.5,0) -- (0,0.5) -- (0.8, 0.5) --  (1.3, 0) --(1.6,0);
\draw (-0.5,0) -- (0,-0.5)-- (0.8, -0.5)--  (1.3, 0);
\filldraw [gray] (-0.5,0) circle (3pt);
\filldraw [gray] (1.3,0) circle (3pt);
\fill[gray] (0.4,0) ellipse (0.2 and 0.6);
\filldraw [black] (-0.2,0.3) circle (2pt);
\filldraw [black] (-0.2,-0.3) circle (2pt);
\filldraw [black] (1,0.3) circle (2pt);
\filldraw [black] (1,-0.3) circle (2pt);
\end{tikzpicture}
\, = \, &
\begin{tikzpicture}
\draw (-0.6,0) -- (-0.5,0);
\draw (0.5,0) -- (0.6,0);
\draw (0, 0.5) -- (0, -0.5);
\draw (0.5,0) arc (0:-180:0.5);
\draw (0.5,0) arc (0:180:0.5);
\end{tikzpicture}
\, + \, 2
\, 
\begin{tikzpicture}
\draw (-0.6,0) -- (-0.5,0);
\draw (0.5,0) -- (0.6,0);
\draw (0.15, 0.475) -- (0.15, -0.475);
\draw (-0.15, 0.475) -- (-0.15, -0.475);
\draw (0.5,0) arc (0:-180:0.5);
\draw (0.5,0) arc (0:180:0.5);
\end{tikzpicture}
\, + \, \phantom{a} 
\,
\begin{tikzpicture}
\draw (-0.6,0) -- (-0.5,0);
\draw (0.5,0) -- (0.6,0);
\draw (-0.35,-0.355) -- (0.35, 0.355);
\filldraw [white] (0,0) circle (3pt);
\draw (-0.35,0.355) -- (0.35, -0.355);
\draw (0.5,0) arc (0:-180:0.5);
\draw (0.5,0) arc (0:180:0.5);
\end{tikzpicture}
+ \cdots,
\end{align*}
but the combination of the two restore the proper count.

\subsection{The problem of overlapping divergences}
A Feynman diagram $\Gamma$ could have divergent evaluation despite a negative superficial degree of divergence $\omega(\Gamma)$. 
This happens whenever a subgraph $\gamma\subset\Gamma$ is divergent, thus with $\omega(\gamma)\geq 0$. 
It was the main contribution of Bogolubov~\cite{BoPa57} to show that a
recursive subtraction of counterterms could suppress such divergences.  Alternatively, in the context
of \SDe, we make use of renormalised vertices.

But how do we generalize this approach to the case of overlapping divergences?
Take as an example 
\begin{equation}  \label{overlap}
\begin{tikzpicture}[scale=0.5]
\draw (-2,0) -- (-1,0); \draw (1,0)--(2,0);
\draw (0,0) circle (1);
\draw (0,1) -- (0,-1);
\draw [green!70!black] (-1.3,1.25) -- (0.5,1.25) -- (0.5,-1.25) -- (-1.3,-1.25) -- (-1.3,1.25);
\end{tikzpicture}
\quad
\text{or}
\quad
\begin{tikzpicture}[scale=0.5]
\draw (-2,0) -- (-1,0); \draw (1,0)--(2,0);
\draw (0,0) circle (1);
\draw (0,1) -- (0,-1);
\draw [green!50!black] (1.3,1.25) -- (-0.5,1.25) -- (-0.5,-1.25) -- (1.3,-1.25) -- (1.3,1.25);
\end{tikzpicture},
\end{equation}
with its two possible subdivergence structures, which overlap.  We would prefer a simple tree structure,
clearly separating all primitive contributions in an expressions while avoiding double counting. 
If we look at equation~\eqref{SDeq_compl}, the first correction is fine, with a simple loop integration if we
replace all vertices with their renormalized value, but the second one has visibly overlapping
subdivergences, and the compensations between the two show that each one should be more difficult to
evaluate than it seems.

The rather old solution, first proposed by Ward~\cite{Wa51}, is to derive with respect to the external
momentum.  The remark that the derivative of a propagator has better ultraviolet behavior is quite old
and is at the base of the Bogolubov method~\cite{BoPa57}. Once the subdivergences are properly dealt
with, a sufficient number of derivations makes the diagram convergent and its renormalized evaluation is
obtained through integration of this finite result.  The integration constants correspond exactly to the
parameters of the Lagrangian and their arbitrariness corresponds to
the choice of renormalisation scheme.

For us, it will be sufficient that the inclusion of the derivative of a
propagator in a three-point function makes it superficially convergent.  If we mark the propagator with
a derivation by a heavy line, one see that now the following diagram has only one possible subdivergence:
\begin{equation}
\begin{tikzpicture}[scale=0.5]
\draw (-2,0) -- (-1,0) arc (180:0:1) --(2,0);
\draw (-1,0) arc (-180:-90:1);
\draw [ultra thick] (1,0) arc (0:-90:1);
\draw (0,1) -- (0,-1);
\draw [green!70!black] (-1.3,1.25) -- (0.5,1.25) -- (0.5,-1.25) -- (-1.3,-1.25) -- (-1.3,1.25);
\end{tikzpicture}
\end{equation}
The highlighted line splits the subdivergences, in the sense that there are those coming before it, and
those coming after it in the flow of momentum through the diagram. For the derivative of the two-point
function, there are many possible primitive diagrams beyond the simple one loop one. Here are some examples
\begin{equation}\label{primitives}
\begin{tikzpicture}
\draw (-0.8,0) -- (-0.5,0);
\draw (0.5,0) -- (0.8,0);
\draw [ultra thick](-140:0.5) arc (-140:-40:0.5);
\draw (220:0.5) arc (220:-40:0.5);
\draw (220:0.5) -- (0,0) -- (-40:0.5);
\draw (0,0) -- (0,0.5);
\filldraw [gray] (-0.5,0) circle (1.5pt);
\filldraw [gray] (0.5,0) circle (1.5pt);
\filldraw [gray] (0,0) circle (1.5pt);
\filldraw [gray] (-140:0.5) circle (1.5pt);
\filldraw [gray] (-40:0.5) circle (1.5pt);
\filldraw [gray] (0, 0.5) circle (1.5pt);
\end{tikzpicture}
\quad
\begin{tikzpicture}
\draw (-0.8,0) -- (-0.5,0);
\draw (0.5,0) -- (0.8,0);
\draw (0,0.5) -- (0,0);
\draw [ultra thick](0,0) -- (-0.4, -0.3);
\draw (0,0) -- (0.4, -0.3);
\draw (0.5,0) arc (0:-180:0.5);
\draw (0.5,0) arc (0:180:0.5);
\filldraw [gray] (-0.5,0) circle (1.5pt);
\filldraw [gray] (0.5,0) circle (1.5pt);
\filldraw [gray] (0,0) circle (1.5pt);
\filldraw [gray] (0.4, -0.3) circle (1.5pt);
\filldraw [gray] (-0.4, -0.3) circle (1.5pt);
\filldraw [gray] (0, 0.5) circle (1.5pt);
\end{tikzpicture}
\quad
\begin{tikzpicture}[scale=0.5]
\draw (-2.5,0) -- (-2,0)-- (-1,1)  -- (1,1) -- (2,0)--(2.5,0);
\draw (-2,0) -- (-1,-1) -- (-0.2,-0.2);
\draw (0.2,.2)--(1,1);
\draw (-1,1) -- (1,-1) -- (2,0);
\draw [ultra thick](-1,-1) -- (1,-1);
\filldraw [gray] (-1,1) circle (3pt);
\filldraw [gray] (1,1) circle (3pt);
\filldraw [gray] (-2,0) circle (3pt);
\filldraw [gray] (2,0) circle (3pt);
\filldraw [gray] (-1,-1) circle (3pt);
\filldraw [gray] (1,-1) circle (3pt);
\end{tikzpicture}
\quad \cdot
\end{equation}
It remains however to know how these diagrams contribute to the derivative of the two-point
function.

The first step is to
derive the equation~\eqref{propagator} for the two-point function:
\begin{equation}\label{devpropagator}
\begin{tikzpicture}
\draw (-0.5,0) -- (0.5,0);
\filldraw [gray] (-0.1,-0.1) -- (-0.1,0.1) -- (0.1,0.1) -- (0.1,-0.1);
\end{tikzpicture}
^\mu
:= 
\partial^\mu  
\begin{tikzpicture}
\draw (-0.5,0) -- (0.5,0);
\filldraw [gray] (0,0) circle (3pt);
\end{tikzpicture}
=
\begin{tikzpicture}
\draw (-0.5,0) -- (0.5,0);
\end{tikzpicture}
^\mu
- \frac{1}{2}\;\;
\begin{tikzpicture}
\draw (-0.9,0) -- (-0.5,0);
\draw (0.5,0) -- (0.7,0);
\draw (0.5,0) arc (0:-180:0.5);
\draw (0.5,0) arc (0:180:0.5);
\filldraw [gray] (-0.6,-0.1) -- (-0.6, 0.1) -- (-0.4, 0.1) -- (-0.4,-0.1);
\filldraw [black] (0, 0.5) circle (2pt);
\filldraw [black] (0,-0.5) circle (2pt);
\end{tikzpicture}
\quad
- \frac{1}{2}\;\;
\begin{tikzpicture}
\draw (-0.8,0) -- (-0.5,0);
\draw (0.5,0) -- (0.8,0);
\draw (0.5,0) arc (0:-180:0.5);
\draw (0.5,0) arc (0:180:0.5);
\filldraw [gray] (-0.5,0) circle (3pt);
\filldraw [black] (0, 0.5) circle (2pt);
\filldraw [white] (-0.1,-0.4) -- (0.1,-0.4) -- (0.1,-0.6) --(-0.1,-0.6) -- cycle;
\filldraw[black] (-0.1,-0.4) -- (0.1,-0.4) -- (0.1,-0.6) --(-0.1,-0.6) -- cycle;
\end{tikzpicture}
\end{equation}
This equation involves the derivative of the vertex, for which one obtain an equation by deriving the
equation~\eqref{vertex}:
\begin{equation}\label{devvertex}
\begin{tikzpicture}
\draw (-0.5,0) -- (0,0);
\draw (0,0) -- (0.5,0.5);
\draw (0,0) -- (0.5,-0.5);
\filldraw [gray] (-0.1,-0.1) -- (-0.1,0.1) -- (0.1,0.1) -- (0.1,-0.1);
\end{tikzpicture}
^\mu
=
\begin{tikzpicture}
\draw (-0.8,0) -- (-0.5,0) -- (0,0.5) -- (0.8, 0.5);
\draw (-0.5,0) -- (0,-0.5)-- (0.8, -0.5);
\filldraw [gray] (-0.6,-0.1) -- (-0.6, 0.1) -- (-0.4, 0.1) -- (-0.4,-0.1);
\fill[gray] (0.4,0) ellipse (0.2 and 0.6);
\filldraw [black] (-0.2,0.3) circle (2pt);
\filldraw [black] (-0.2,-0.3) circle (2pt);
\end{tikzpicture}
+
\begin{tikzpicture}
\draw (-0.8,0) -- (-0.5,0) -- (0,0.5) -- (0.8, 0.5);
\draw (-0.5,0) -- (0,-0.5)-- (0.8, -0.5);
\filldraw [gray] (-0.5,0) circle (3pt);
\fill[gray] (0.4,0) ellipse (0.2 and 0.6);
\filldraw [black] (-0.2,0.3) circle (2pt);
\filldraw [white] (-0.3,-0.4) -- (-0.1,-0.4) -- (-0.1,-0.2) -- (-0.3,-0.2) -- cycle;
\filldraw[black] (-0.3,-0.4) -- (-0.1,-0.4) -- (-0.1,-0.2) -- (-0.3,-0.2) -- cycle;
\end{tikzpicture}
+
\begin{tikzpicture}
\draw (-0.8,0) -- (-0.5,0) -- (0,0.5) -- (0.8, 0.5);
\draw (-0.5,0) -- (0,-0.5)-- (0.8, -0.5);
\filldraw [gray] (-0.5,0) circle (3pt);
\filldraw [gray] (0.2,0.5) -- (0.6, 0.5) -- (0.6, -0.5) -- (0.2,-0.5);
\filldraw [black] (-0.2,0.3) circle (2pt);
\filldraw [black] (-0.2,-0.3) circle (2pt);
\end{tikzpicture}
\end{equation}
where the squares represent the derived objects and the round ones the original ones.
We can now use equation~\eqref{vertex} to re-express the right bare vertices  in the derived propagator~\eqref{devpropagator} and obtain 
\begin{equation}
\begin{tikzpicture}
\draw (-0.5,0) -- (0.5,0);
\filldraw [gray] (-0.1,-0.1) -- (-0.1,0.1) -- (0.1,0.1) -- (0.1,-0.1);
\end{tikzpicture}
^\mu
=
\begin{tikzpicture}
\draw (-0.5,0) -- (0.5,0);
\end{tikzpicture}
^\mu
- \frac{1}{2}\;
\Bigg(
\begin{tikzpicture}
\draw (-0.9,0) -- (-0.5,0);
\draw (0.5,0) -- (0.7,0);
\draw (0.5,0) arc (0:-180:0.5);
\draw (0.5,0) arc (0:180:0.5);
\filldraw [gray] (-0.6,-0.1) -- (-0.6, 0.1) -- (-0.4, 0.1) -- (-0.4,-0.1);
\filldraw [gray] (0.5,0) circle (3pt);
\filldraw [black] (0, 0.5) circle (2pt);
\filldraw [black] (0,-0.5) circle (2pt);
\end{tikzpicture}
\;
-
\;
\begin{tikzpicture}
\draw (-0.8,0) -- (-0.5,0) -- (0,0.5) -- (0.8, 0.5) --  (1.3, 0) --(1.6,0);
\draw (-0.5,0) -- (0,-0.5)-- (0.8, -0.5)--  (1.3, 0);
\filldraw [gray] (-0.6,-0.1) -- (-0.6, 0.1) -- (-0.4, 0.1) -- (-0.4,-0.1);
\filldraw [gray] (1.3,0) circle (3pt);
\fill[gray] (0.4,0) ellipse (0.2 and 0.6);
\filldraw [black] (-0.2,0.3) circle (2pt);
\filldraw [black] (-0.2,-0.3) circle (2pt);
\filldraw [black] (1,0.3) circle (2pt);
\filldraw [black] (1,-0.3) circle (2pt);
\end{tikzpicture}
\;
+
\;
\begin{tikzpicture}
\draw (-0.8,0) -- (-0.5,0);
\draw (0.5,0) -- (0.8,0);
\draw (0.5,0) arc (0:-180:0.5);
\draw (0.5,0) arc (0:180:0.5);
\filldraw [gray] (-0.5,0) circle (3pt);
\filldraw [gray] (0.5,0) circle (3pt);
\filldraw [black] (0, 0.5) circle (2pt);
\filldraw [white] (-0.1,-0.4) -- (0.1,-0.4) -- (0.1,-0.6) --(-0.1,-0.6) -- cycle;
\filldraw[black] (-0.1,-0.4) -- (0.1,-0.4) -- (0.1,-0.6) --(-0.1,-0.6)--cycle;
\end{tikzpicture}
\;
-
\;
\begin{tikzpicture}
\draw (-0.8,0) -- (-0.5,0) -- (0,0.5) -- (0.8, 0.5) --  (1.3, 0) --(1.6,0);
\draw (-0.5,0) -- (0,-0.5)-- (0.8, -0.5)--  (1.3, 0);
\filldraw [gray] (-0.5,0) circle (3pt);
\filldraw [gray] (1.3,0) circle (3pt);
\fill[gray] (0.4,0) ellipse (0.2 and 0.6);
\filldraw [black] (-0.2,0.3) circle (2pt);
\filldraw [white] (-0.3,-0.4) -- (-0.1,-0.4) -- (-0.1,-0.2) -- (-0.3,-0.2) -- cycle;
\filldraw[black] (-0.3,-0.4) -- (-0.1,-0.4) -- (-0.1,-0.2) -- (-0.3,-0.2)--cycle;
\filldraw [black] (1,0.3) circle (2pt);
\filldraw [black] (1,-0.3) circle (2pt);
\end{tikzpicture}
\Bigg)
\end{equation}
which ultimately, by use of equation~\eqref{devvertex} in the first term of the parenthesis, simplifies to:
 \begin{equation}\label{SDe_K}
\begin{tikzpicture}
\draw (-0.5,0) -- (0.5,0);
\filldraw [gray] (-0.1,-0.1) -- (-0.1,0.1) -- (0.1,0.1) -- (0.1,-0.1);
\end{tikzpicture}
^\mu
= 
\begin{tikzpicture}
\draw (-0.5,0) -- (0.5,0);
\end{tikzpicture}
^\mu
\;\;
- \frac{1}{2}\;\;
\begin{tikzpicture}
\draw (-0.8,0) -- (-0.5,0);
\draw (0.5,0) -- (0.8,0);
\draw (0.5,0) arc (0:-180:0.5);
\draw (0.5,0) arc (0:180:0.5);
\filldraw [gray] (-0.5,0) circle (3pt);
\filldraw [gray] (0.5,0) circle (3pt);
\filldraw [black] (0, 0.5) circle (2pt);
\filldraw [white] (-0.1,-0.4) -- (0.1,-0.4) -- (0.1,-0.6) --(-0.1,-0.6)--cycle;
\filldraw[black] (-0.1,-0.4) -- (0.1,-0.4) -- (0.1,-0.6) --(-0.1,-0.6)--cycle;
\end{tikzpicture}
\;\;
 - \frac{1}{2}\;\;
 \begin{tikzpicture}
 \draw (-0.8,0) -- (-0.5,0) -- (0,0.5) -- (0.8, 0.5) --  (1.3, 0) --(1.6,0);
 \draw (-0.5,0) -- (0,-0.5)-- (0.8, -0.5)--  (1.3, 0);
 \filldraw [gray] (-0.5,0) circle (3pt);
 \filldraw [gray] (1.3,0) circle (3pt);
 \filldraw [gray] (0.2,0.5) -- (0.6, 0.5) -- (0.6, -0.5) -- (0.2,-0.5);
 \filldraw [black] (-0.2,0.3) circle (2pt);
 \filldraw [black] (-0.2,-0.3) circle (2pt);
 \filldraw [black] (1,0.3) circle (2pt);
 \filldraw [black] (1,-0.3) circle (2pt);
 \end{tikzpicture}
\quad .
 \end{equation}
If we expand the last term by deriving the expression for the four-point kernel~\eqref{4ptfunction} we have
\begin{equation}
 \begin{tikzpicture}
 \draw (-0.8,0) -- (-0.5,0) -- (0,0.5) -- (0.8, 0.5) --  (1.3, 0) --(1.6,0);
 \draw (-0.5,0) -- (0,-0.5)-- (0.8, -0.5)--  (1.3, 0);
 \filldraw [gray] (-0.5,0) circle (3pt);
 \filldraw [gray] (1.3,0) circle (3pt);
 \filldraw [gray] (0.2,0.5) -- (0.6, 0.5) -- (0.6, -0.5) -- (0.2,-0.5);
 \filldraw [black] (-0.2,0.3) circle (2pt);
 \filldraw [black] (-0.2,-0.3) circle (2pt);
 \filldraw [black] (1,0.3) circle (2pt);
 \filldraw [black] (1,-0.3) circle (2pt);
 \end{tikzpicture}
 =
 \begin{tikzpicture}
	\draw (-0.8,0) -- (-0.5,0);
	\draw (0.5,0) -- (0.8,0);
	\draw (0, 0.5) -- (0, -0.5);
	\draw (0.5,0) arc (0:-180:0.5);
	\draw (0.5,0) arc (0:180:0.5);
	\filldraw [gray] (-0.1,-0.4) -- (0.1,-0.4) -- (0.1,-0.6) --(-0.1,-0.6);
	\filldraw [gray] (-0.5,0) circle (3pt);
	\filldraw [gray] (0.5,0) circle (3pt);
	\filldraw [gray] (0,0.5) circle (3pt);
\end{tikzpicture}
 \, + \frac{1}{2}\,
\begin{tikzpicture}
	\draw (-0.8,0) -- (-0.5,0);
	\draw (0.5,0) -- (0.8,0);
	\draw (-0.35,-0.355) -- (0.35, 0.355);
	\filldraw [white] (0,0) circle (3pt);
	\draw (-0.35,0.355) -- (0.35, -0.355);
	\draw (0.5,0) arc (0:-180:0.5);
	\draw (0.5,0) arc (0:180:0.5);
	\filldraw [gray] (-0.45,-0.255)-- (-0.25,-0.255) -- (-0.25,-0.455) -- (-0.45,-0.455);
	\filldraw [gray] (-0.5,0) circle (3pt);
	\filldraw [gray] (0.5,0) circle (3pt);
	\filldraw [gray] (0.35, 0.355) circle (3pt);
	\filldraw [gray] (-0.35,0.355) circle (3pt);
	\filldraw [gray] (0.35, -0.355) circle (3pt);
\end{tikzpicture}
 \, +\frac{1}{2} \,
\begin{tikzpicture}
\draw (-0.8,0) -- (-0.5,0);
\draw (0.5,0) -- (0.8,0);
\draw (-0.35,-0.355) -- (0.35, 0.355);
\filldraw [white] (0,0) circle (3pt);
\draw (-0.35,0.355) -- (0.35, -0.355);
\draw (0.5,0) arc (0:-180:0.5);
\draw (0.5,0) arc (0:180:0.5);
\filldraw [white] (-0.1,-0.4) -- (0.1,-0.4) -- (0.1,-0.6) --(-0.1,-0.6);
\filldraw [gray] (-0.1,-0.4) -- (0.1,-0.4) -- (0.1,-0.6) --(-0.1,-0.6)--cycle;
\filldraw [gray] (-0.5,0) circle (3pt);
\filldraw [gray] (0.5,0) circle (3pt);
\filldraw [gray] (0.35, 0.355) circle (3pt);
\filldraw [gray] (-0.35,0.355) circle (3pt);
\filldraw [gray] (0.35, -0.355) circle (3pt);
\filldraw [gray] (-0.35, -0.355) circle (3pt);
\end{tikzpicture}
 \, + \frac{1}{2}\,
\begin{tikzpicture}
	\draw (-0.8,0) -- (-0.5,0);
	\draw (0.5,0) -- (0.8,0);
	\draw (-0.35,-0.355) -- (0.35, 0.355);
	\filldraw [white] (0,0) circle (3pt);
	\draw (-0.35,0.355) -- (0.35, -0.355);
	\draw (0.5,0) arc (0:-180:0.5);
	\draw (0.5,0) arc (0:180:0.5);
	\filldraw [gray] (0.45,-0.255)-- (0.25,-0.255) -- (0.25,-0.455) -- (0.45,-0.455);
	\filldraw [gray] (-0.5,0) circle (3pt);
	\filldraw [gray] (0.5,0) circle (3pt);
	\filldraw [gray] (0.35, 0.355) circle (3pt);
	\filldraw [gray] (-0.35,0.355) circle (3pt);
	\filldraw [gray] (-0.35, -0.355) circle (3pt);
\end{tikzpicture}
\, + \ldots \,
 \end{equation}
Whenever we have a derivative of a vertex, we should make use of equation~\eqref{devvertex} to obtain a
more explicit value, with the remaining parts involving derivatives of the vertex or the 4-point kernel having
to be recursively expanded.  The final objects we want to evaluate should have the derivative only on
one propagator.  We could in this way obtain certain of the diagrams in
equation~\eqref{primitives} with determined weights.  However, since the first contributions
beyond the one-loop one appear at three loop order, we will not need them in this work, but should keep
in mind that the full solution will require an infinite set of primitives.

This concludes the determination of the equations that we will use.  As in many cases in quantum field
theory, it is not clear whether this derivation can be made rigorous, since we should at least put some
regularization.  In a sense, this is not really important if we can show that the solutions of these
equations have the properties of the Green functions of a quantum field theory.  

\subsection{Renormalised propagators} 
As renormalised quantities, all our $n$-point functions depend on the coupling constant $g$ and a
renormalisation momentum scale $\mu$.
The two point function $P$ has more natural variables: $a:=g^2/ \left( 2\sqrt{\pi} \right)^D$ since all
the graphs involved have an even number of vertices, and $L := \log (p^2/\mu^2)$ since the general
solution to a Callan--Symanzik equation is a series in $g$ with coefficients that are polynomials in $L$
(see for instance~\cite{Collins}, chap 5). We can represent $P(a,L)$ as 
\begin{equation}
	P(a,L) = \frac 1{p^2} G(a,L) := \frac 1{p^2}
	\sum_{n=0} \frac{\gamma_n(a)}{n!} L^n   .
\end{equation}
The free propagator $1/p^2$ has been factored out and the series $G(a,L)$ contains all the corrections
due to the renormalisation process. In this representation, the first two coefficients of the series in
\(L\) have special significance: $\gamma_0 = 1$, reflecting the renormalisation condition that the
propagator is unchanged at the renormalisation point $L=0$, and $\gamma_1=\gamma$ the anomalous
dimension of the field.

Noting that $L^n = {\partial^n_x} e^{xL}|_{x=0}$, we can describe the series \(G(a,L)\) as a
pseudodifferential operator obtained by replacing $L$ by $\mathlarger{\partial_x}$ in the series, acting
on \(e^{xL}\):
\begin{equation}
G(a,L) = \left. G(a,{\partial_x}) \left(p^2/\mu^2 \right)^x\right|_{x=0}.
\end{equation}
From now on, we will not indicate the evaluation at $x=0$ but it remains implied.

The derivation $\D$ with respect to $L$ will be important for us:
\begin{equation}
\D G(a,L) = \sum_{n=0} \frac{\gamma_{n+1}(a)}{n!} L^n.
\end{equation}
In the context of the corresponding
pseudodifferential operators, the iterations of $\D$ have a simple expression
\begin{equation}\label{xmGm}
G(a, \partial_x) \left(x^m \cdot f \right) = (\D^m G)(a, \partial_x) \, f
\end{equation}
as we can easily show by direct inspection
\begin{equation}
	G(a, \partial_x) \left(x^m \cdot f \right) = \sum_{n=0} \frac{\gamma_n(a)}{n!}  \,\partial_x^n
	\left( x^m \cdot f \right) = \sum_{n=m} \frac{\gamma_n(a)}{(n-m)!}  \,\partial_x^{n-m} f 
\end{equation}
using that $\partial_x^k x^m= m!\delta_{km}$ when evaluated at $0$ and the binomial expansion of the
higher derivatives of a product.

The Ward--Schwinger--Dyson equations depend on other quantities for which we will also need convenient
expression. First we have the relation between the $2$-point function (the propagator) and the
corresponding vertex function (the 1PI two-point function)
\begin{equation}\label{PGamma2}
P(a,L) \cdot \Gamma_2(a,L) = 1,
\end{equation}
where the $1$ comes from the fact that we are working in Euclidean space.  The first few terms of
$\Gamma_2(a,L)$ are given by 
\begin{equation}\label{Gamma2development}
	\Gamma_2(a,L) = p^2 \left( 1-\gamma_1 L + \left( 2\gamma_1^2- \gamma_2 \right) L^2/2 + O(L^3)
	\right) .
\end{equation}
We will need the derivatives with respect to \(p_\nu\) of both \(P\) and \(\Gamma_2\).  This derivation
will be abbreviated to \(\partial^\nu\) in many cases. We have 
\begin{equation}\label{refKnu}
K^\nu (a,L) := \frac{\partial}{\partial p_\nu} P(a,L) = \frac{2p^\nu}{(p^2)^2} \left( \D G(a,L) - G(a,L) \right) 
\end{equation}
since $\partial^\nu L = 2p^\nu/p^2$. This can be expressed in terms of the \(\gamma_n\):
\begin{equation}\label{defKnu}
K^\nu (a,L) = \frac{2p^\nu}{(p^2)^2} \mathlarger{\sum}_{n=0} \frac{\gamma_{n+1}(a)-\gamma_n(a)}{n!} L^n.
\end{equation}
Likewise we define the derivative of the $\Gamma_2$ function,  
\begin{equation}\label{Cdevelopement}
C^\nu (a,L) := \partial^\nu \Gamma_2 (a,L) = 2p^\nu \left( \D G^{-1}(a,L) + G^{-1}(a,L) \right) 
\end{equation}
with its coefficients obtained by adding two subsequent coefficients in the development of $G^{-1}$.
The derivation of equation~\eqref{PGamma2} gives a relation between $K^\nu$ and $C^\nu$  
\begin{equation}
K^\nu (a,L) = - P(a,L) \cdot C^\nu (a,L)\cdot P(a,L) 
\end{equation}
which graphically becomes 
\begin{equation}
\begin{tikzpicture}
	\draw (-0.5,0) -- (0.5,0);
	\filldraw [white] (-0.1,0.1) -- (0.1,0.1) -- (0.1,-0.1) --(-0.1,-0.1)--cycle;
	\filldraw[black] (-0.1,-0.1) -- (-0.1,0.1) -- (0.1,0.1) -- (0.1,-0.1)--cycle;
	\node at (0.3,0.2) {$\nu$};
\end{tikzpicture}
=  - \;
\begin{tikzpicture}
	\draw (-1.5,0) -- (1.5,0);
	\filldraw [black] (-1,0) circle (3pt);
	\filldraw [black] (1,0) circle (3pt);
	\filldraw [gray] (-0.1,-0.1) -- (-0.1,0.1) -- (0.1,0.1) -- (0.1,-0.1);
	\node at (0.3,0.2) {$\nu$};
\end{tikzpicture}
\end{equation}

\subsection{Infrared rearrangement}\label{IR_rearrangement}

We still need a way to treat the renormalised vertex function $\Gamma_3$. The full three-point function
depends on all the possible scalar products  of the external momenta minus the delta function condition,
so in our case $3$ variables. Their functional dependence on these kinematic invariants are unknown at
higher loop order, but even the known cases do not easily allow for the evaluation of diagrams including
vertices with the full dependence on all invariants.

A guide for a solution is that the renormalisation functions only depend on the three-point function evaluated
at a multiple of the renormalisation condition.  We will therefore try to only use this special case of
the three-point function, which will have the same kind of functional dependence as the two-point one.
This will use the trick called infrared rearrangement, where one transforms a diagram in one with the
same subdivergences, but with simpler evaluation.  A nice characteristics of this six-dimensional theory
is that we do not introduce infrared divergences in the process.

In the definition of the vertex in equation~\eqref{vertex} we consider the case where one incoming momentum is
set to zero. In Euclidean space, the null momentum condition is equivalent to the one of a zero norm for
the momentum and imposes that the invariants associated to the two other inputs of the vertex diagram are equal. 
A dotted line signals the zero momentum input.
\begin{equation}\label{vertex0}
\begin{tikzpicture}
\draw[dotted, thick] (-0.5,0) -- (0,0);
\draw (0,0) -- (0.5,0.5);
\draw (0,0) -- (0.5,-0.5);
\filldraw [gray] (0,0) circle (3pt);
\end{tikzpicture}
=
\begin{tikzpicture}
\draw [dotted,thick] (-0.5,0) -- (0,0);
\draw (0,0) -- (0.5,0.5);
\draw (0,0) -- (0.5,-0.5);
\end{tikzpicture}
+\;\;
\begin{tikzpicture}
\draw[dotted, thick] (-0.8,0) -- (-0.5,0);
\draw  (-0.5,0) -- (0,0.5) -- (0.8, 0.5);
\draw (-0.5,0) -- (0,-0.5)-- (0.8, -0.5);
\filldraw [gray] (-0.5,0) circle (3pt);
\fill[gray] (0.4,0) ellipse (0.2 and 0.6);
\filldraw [black] (0,0.5) circle (2pt);
\filldraw [black] (0,-0.5) circle (2pt);
\end{tikzpicture}
\end{equation}
For the bare vertex, which does not depend on incoming momenta, the dotted line does not change its evaluation.
The last term is well defined since in 6~dimensions, the product of two propagators with the same momentum is
not infrared divergent.  This single-scale configuration of momenta is the only one for which we will evaluate
the vertex.

However, in this very equation, the
vertices included in the four-point kernel have generic configurations of the inputs.  We therefore want to
reexpress all diagrams using only this single scale approximation of the vertex, by a process known as infrared
rearrangement.  This is made by joining two of the external legs of the vertex on one of its inputs while letting
an other one of the inputs without any connection to the rest of the diagram and therefore with a zero momentum
entry.  Graphically, this is represented by two lines connecting to a single point of the boundary of a blob,
while an interrupted dotted lines represent the unused third input of the vertex.  This can be further simplified
by using the following convention 
\begin{equation}
\begin{tikzpicture}
\draw (-0.5,0) -- (0,0);
\draw (0,0) -- (0.5,0.5);
\draw (0,0) -- (0.5,-0.5);
\filldraw [gray] (0,0) -- (-0.05,0.3) -- (0.2, 0.2) ;
\end{tikzpicture}
:=
\begin{tikzpicture}
\draw (-0.5,0) -- (0,0);
\draw (0,0) -- (0.5,0.5);
\draw (0,0) -- (0.5,-0.5);
\draw [dotted, thick] (-0.2,0.4) -- (0.1,0.1);
\filldraw [gray] (0.1,0.1) circle (3pt);
\end{tikzpicture}
\end{equation}
to denote the one scale object. 
Like the correction to the propagator, it can be characterised by a formal series,
again in~$\log (p^2/\mu^2)$:
\begin{equation}\label{Gamma_3}
\Gamma_3(p;a) := g \; Y(a,L) = g \;\mathlarger{\sum}_{n\geq0} \upsilon_n (a) \frac{L^n}{n!}
\end{equation}
with again the convention that the first term of the series $\upsilon_0$ is set to $1$, ensuring the
renormalisation condition when $p^2=\mu^2$.

Substituting the one scale object in the diagrams allows for their simple evaluation, since only logarithms of
the momenta in some propagators are introduced, but introduces errors which must be controlled and computed. We
therefore have to compute the following difference
\begin{equation*}
\begin{tikzpicture}
\draw (-0.5,0) -- (0,0);
\draw (0,0) -- (0.5,0.5);
\draw (0,0) -- (0.5,-0.5);
\filldraw [gray] (0,0) circle (3pt);
\end{tikzpicture}
\, - \, 
\begin{tikzpicture}
\draw (-0.5,0) -- (0,0);
\draw (0,0) -- (0.5,0.5);
\draw (0,0) -- (0.5,-0.5);
\filldraw [gray] (0,0) -- (-0.05,0.3) -- (0.2, 0.2) ;
\end{tikzpicture}
\end{equation*}
with both terms evaluated using, respectively, equations \eqref{vertex} and~\eqref{vertex0}. The bare vertex is
independent of all momenta and thus the same in both equations so that we obtain:
\begin{equation}\label{deformationvertex}
\begin{tikzpicture}
\draw (-0.5,0) -- (0,0);
\draw (0,0) -- (0.5,0.5);
\draw (0,0) -- (0.5,-0.5);
\filldraw [gray] (0,0) circle (3pt);
\end{tikzpicture}
\, - \, 
\begin{tikzpicture}
\draw (-0.5,0) -- (0,0);
\draw (0,0) -- (0.5,0.5);
\draw (0,0) -- (0.5,-0.5);
\filldraw [gray] (0,0) -- (-0.05,0.3) -- (0.2, 0.2) ;
\end{tikzpicture}
=\;\;
\begin{tikzpicture}
\draw (-0.8,0) -- (-0.5,0);
\draw  (-0.5,0) -- (0,0.5) -- (0.8, 0.5);
\draw (-0.5,0) -- (0,-0.5)-- (0.8, -0.5);
\filldraw [gray] (-0.5,0) circle (3pt);
\fill[gray] (0.4,0) ellipse (0.2 and 0.6);
\filldraw [black] (0,0.5) circle (2pt);
\filldraw [black] (0,-0.5) circle (2pt);
\end{tikzpicture}
-\;\;
\begin{tikzpicture}
\draw (-0.8,-0.4) -- (0,-0.4) -- (0.2, -0.5) -- (0.8, -0.5) ;
\draw (0,0.5) -- (0.8,0.5);
\draw [dotted, thick] (-0.8,0.05) -- (-0.4,0.05);
\draw (0,0.5) arc (90: 270: 0.45) ; 
\fill[gray] (0,0.05) ellipse (0.1 and 0.45);
\filldraw [gray] (-0.45,0.05) circle (2.5pt);
\filldraw [black] (-0.3,-0.25) circle (2pt);
\filldraw [black] (-0.3,0.35) circle (2pt);
\end{tikzpicture}
\end{equation}
The two terms in the right-hand side present the same subdivergences, so that substitution of this difference in
a diagram will produce a primitive composite diagram. It means that this correction will
produce contributions subleading in powers of the log at a given order in the coupling. This is akin to the angle
scale separation introduced in~\cite{BrKr2013}, which was an inspiration for this part of our work. 
A problem which may be feared is that this process may introduce infrared divergences.  However, contrary to the case of \(\phi_4^4\) studied for example in~\cite{KoPa2016}, where a particular choice must be taken to avoid them, infrared divergences never appear in our case. This happens because the sequence of two propagators is not infrared divergent in 6 dimensions.

A particular care should be paid to the last term in equation~\eqref{deformationvertex}, since it seems
to involve a elementary four-particle vertex, when the expansion from
equation~\eqref{4ptfunction} is inserted in it. The two branches must merge into a single input outside of every
expansion of the blobs. This means that in the resulting diagrams, vertices are no longer all three-valent.

We can now go back to equation~\eqref{SDeq_compl} and insert equation~\eqref{deformationvertex} in it to
obtain
\begin{align}\label{prop_corr}
\begin{tikzpicture}
\draw (-0.8,0) -- (-0.5,0);
\draw (0.5,0) -- (0.8,0);
\draw (0.5,0) arc (0:-180:0.5);
\draw (0.5,0) arc (0:180:0.5);
\filldraw [gray] (-0.5,0) circle (3pt);
\filldraw [gray] (0.5,0) circle (3pt);
\filldraw [white] (-0.1,-0.4) -- (0.1,-0.4) -- (0.1,-0.6) --(-0.1,-0.6)--cycle;
\filldraw[black] (-0.1,-0.4) -- (0.1,-0.4) -- (0.1,-0.6) --(-0.1,-0.6)--cycle;
\end{tikzpicture}
\quad
&=
\quad
\begin{tikzpicture}
\draw (-0.8,0) -- (-0.5,0);
\draw (0.5,0) -- (0.8,0);
\draw (-0.5,0) -- (-0.6,0.3)-- (-0.4,0.3);
\draw (0.5,0) -- (0.6,0.3) -- (0.4,0.3);
\fill [gray] (-0.5,0) -- (-0.6,0.3) -- (-0.4,0.3);  
\fill [gray] (0.5,0) -- (0.6,0.3) -- (0.4,0.3);  
\fill [white] (0.5,0) arc (0:180:0.5);
\draw (0.5,0) arc (0:-180:0.5);
\draw (0.5,0) arc (0:180:0.5);
\filldraw [white] (-0.1,-0.4) -- (0.1,-0.4) -- (0.1,-0.6) --(-0.1,-0.6)--cycle;
\filldraw[black] (-0.1,-0.4) -- (0.1,-0.4) -- (0.1,-0.6) --(-0.1,-0.6)--cycle;
\end{tikzpicture}
\quad
+
\quad
2
\quad
\Bigg(
\begin{tikzpicture}
\draw (0.8,0) -- (1,0);
\draw (-1,0) -- (-0.8,0);
\fill [white] (-0.3,0.5) arc (90:270:0.5);
\draw (-0.3,0.5) arc (90:270:0.5);
\fill [gray] (-0.8,0) -- (-0.9,0.3) -- (-0.7,0.3);  
\draw (0.3,0.5) arc (90:-90:0.5);
\draw (-0.3,0.5) -- (0.3,0.5);
\draw (-0.3,-0.5) -- (0.3,-0.5);
\filldraw [white] (-0.3,-0.4) -- (-0.1,-0.4) -- (-0.1,-0.6) --(-0.3,-0.6)--cycle;
\filldraw[black] (-0.3,-0.4) -- (-0.1,-0.4) -- (-0.1,-0.6) --(-0.3,-0.6)--cycle;
\filldraw [gray] (0.8,0) circle (3pt);
\fill[gray] (0.3,0) ellipse (0.2 and 0.6);
\end{tikzpicture}
\quad
-
\enspace
\begin{tikzpicture}
\draw (-1,0) -- (-0.8,0);
\fill [white] (-0.3,0.5) arc (90:270:0.5);
\draw (-0.3,0.5) arc (90:270:0.5);
\filldraw [gray] (0.8,0.2) -- (0.8,-0.1) -- (1,0.05) -- (0.8,0.2); 
\fill [white] (0.4,0.5) arc (90:-90:0.45);
\draw (0.4,0.5) arc (90:-90:0.45);
\draw (-0.3,0.5) -- (0.4,0.5);
\draw (-0.3,-0.5) -- (0.3,-0.5);
\filldraw [white] (-0.3,-0.4) -- (-0.1,-0.4) -- (-0.1,-0.6) --(-0.3,-0.6)--cycle;
\filldraw[black] (-0.3,-0.4) -- (-0.1,-0.4) -- (-0.1,-0.6) --(-0.3,-0.6)--cycle;
\fill [gray] (-0.8,0) -- (-0.9,0.3) -- (-0.7,0.3);  
\draw (0.4,-0.4) -- (1,-0.4);
\draw (0.4,-0.4) arc (0:-90:0.1);
\draw (0.4,0.5) arc (90:-90:0.45);
\fill[gray] (0.4,0.05) ellipse (0.1 and 0.45);
\end{tikzpicture}
\Bigg)
\\
&
\quad
+
\enspace \left(
\begin{tikzpicture}
\draw (-1,0) -- (-0.8,0);
\draw (0.8,0) -- (1,0);
\draw (-0.3,0.5) arc (90:270:0.5);
\draw (0.3,0.5) arc (90:-90:0.5);
\draw (-0.3,0.5) -- (0.3,0.5);
\draw (-0.3,-0.5) -- (0.3,-0.5);
\filldraw [white] (-0.1,-0.4) -- (0.1,-0.4) -- (0.1,-0.6) --(-0.1,-0.6)--cycle;
\filldraw[black] (-0.1,-0.4) -- (0.1,-0.4) -- (0.1,-0.6) --(-0.1,-0.6)--cycle;
\filldraw [gray] (-0.8,0) circle (3pt);
\filldraw [gray] (0.8,0) circle (3pt);
\fill[gray] (0.3,0) ellipse (0.2 and 0.6);
\fill[gray] (-0.3,0) ellipse (0.2 and 0.6);
\end{tikzpicture}
\quad
-
\enspace
\begin{tikzpicture}
\draw (-1,0) -- (-0.8,0);
\fill [white] (-0.3,0.5) arc (90:270:0.5);
\draw (-0.3,0.5) arc (90:270:0.5);
\fill [white] (0.4,0.5) arc (90:-90:0.45);
\draw (0.4,0.5) arc (90:-90:0.45);
\filldraw [gray] (0.84,0.15) -- (0.84,-0.05) -- (1,0.05) -- (0.84,0.15); 
\draw (-0.3,0.5) -- (0.4,0.5);
\draw (-0.3,-0.5) -- (0.3,-0.5);
\filldraw [white] (0,-0.4) -- (0.2,-0.4) -- (0.2,-0.6) --(0,-0.6)--cycle;
\filldraw[black] (0,-0.4) -- (0.2,-0.4) -- (0.2,-0.6) --(0,-0.6)--cycle;
\fill [gray] (-0.8,0) -- (-0.9,0.3) -- (-0.7,0.3);  
\draw (0.4,-0.4) -- (1,-0.4);
\draw (0.4,-0.4) arc (0:-90:0.1);
\fill[gray] (0.4,0.05) ellipse (0.1 and 0.45);
\fill[gray] (-0.3,0) ellipse (0.2 and 0.6);
\end{tikzpicture}
\quad
-
\enspace
\begin{tikzpicture}
\draw (1,0) -- (0.8,0);
\fill [white] (0.3,0.5) arc (90:-90:0.5);
\draw (0.3,0.5) arc (90:-90:0.5);
\fill [white] (-0.4,0.5) arc (90:270:0.45);
\draw (-0.4,0.5) arc (90:270:0.45);
\filldraw [gray] (-0.84,0.15) -- (-0.84,-0.05) -- (-1,0.05) -- (-0.84,0.15); 
\draw (0.3,0.5) -- (-0.4,0.5);
\draw (0.3,-0.5) -- (-0.3,-0.5);
\filldraw [white] (0,-0.4) -- (-0.2,-0.4) -- (-0.2,-0.6) --(0,-0.6)--cycle;
\filldraw[black] (0,-0.4) -- (-0.2,-0.4) -- (-0.2,-0.6) --(0,-0.6)--cycle;
\fill [gray] (0.8,0) -- (0.9,0.3) -- (0.7,0.3);  
\draw (-0.4,-0.4) -- (-1,-0.4);
\draw (-0.4,-0.4) arc (180:270:0.1);
\fill[gray] (-0.4,0.05) ellipse (0.1 and 0.45);
\fill[gray] (0.3,0) ellipse (0.2 and 0.6);
\end{tikzpicture}
\quad
+
\enspace
\begin{tikzpicture}
\fill [white] (0.4,0.5) arc (90:-90:0.45);
\draw (0.4,0.5) arc (90:-90:0.45);
\filldraw [gray] (0.84,0.15) -- (0.84,-0.05) -- (1,0.05) -- (0.84,0.15); 
\fill [white] (-0.4,0.5) arc (90:270:0.45);
\draw (-0.4,0.5) arc (90:270:0.45);
\filldraw [gray] (-0.84,0.15) -- (-0.84,-0.05) -- (-1,0.05) -- (-0.84,0.15); 
\draw (0.4,0.5) -- (-0.4,0.5);
\draw (0.3,-0.5) -- (-0.3,-0.5);
\filldraw [white] (0.1,-0.4) -- (-0.1,-0.4) -- (-0.1,-0.6) -- (0.1,-0.6)--cycle;
\filldraw[black] (0.1,-0.4) -- (-0.1,-0.4) -- (-0.1,-0.6) -- (0.1,-0.6)--cycle;
\draw (-0.4,-0.4) -- (-1,-0.4);
\draw (-0.4,-0.4) arc (180:270:0.1);
\fill[gray] (-0.4,0.05) ellipse (0.1 and 0.45);
\fill[gray] (0.4,0.05) ellipse (0.1 and 0.45);
\draw (0.4,-0.4) -- (1,-0.4);
\draw (0.4,-0.4) arc (0:-90:0.1);
\end{tikzpicture}
\right)
\nonumber
\end{align}
where all propagators are supposed to be full propagators, even if we do not mark them by black circles.
We see that the first term will be the dominant one in the expansion in $a$. The next parenthesis (with
a factor of two indicating that you have the drawn diagrams plus their mirror images) will give a
correction starting at order $a^2$ that will be addressed in section~\ref{IRcorrections}. The last group
start contributing at order $a^3$ and is neglected in this work.  We remark that in all these groupings,
subdivergences are compensated between the different members, so that they behave as primitive
divergences.  We will not try to give an all order ansatz for the generation of all the primitive terms
which will be generated by the recursive application of equation~\eqref{deformationvertex} in the
diagrams.  We suppose that they could be generated by a combinatorial
\SDe\ for the full three-point functions followed by what should look akin to a renormalization
where the counterterms are obtained by moving one of the exterior link of the subdiagram to make it
single scale.

At each stage of our approximation scheme, we only deal with single scale versions of the vertices, with
a logarithmic dependence on one the momentum entering it.  This means that
in our diagrams, the vertices do not introduce any new analytic difficulty, since they just add logarithmic
factors similar to the ones coming from one of the neighbouring lines.
New kind of dressed propagators will appear, like 
\begin{equation}\label{propstrani}
\begin{tikzpicture}
  \draw (0,0) -- (2,0);
  \fill [gray] (0,0) -- (0.25,0.4) -- (0.5,0);  
    \filldraw [black] (1,0) circle (3pt);
 \end{tikzpicture}
 \quad
 \begin{tikzpicture}
  \draw (0,0) -- (2,0);
  \fill [gray] (0.75,0) -- (1,0.4) -- (1.25,0);  
  \filldraw [black] (0.4,0) circle (3pt);
  \filldraw [black] (1.6,0) circle (3pt);
 \end{tikzpicture}
 \quad
 \begin{tikzpicture}
  \draw (0,0) -- (2,0);
  \fill [gray] (0,0) -- (0.25,0.4) -- (0.5,0);  
  \fill [gray] (1.5,0) -- (1.75,0.4) -- (2,0);  
    \filldraw [black] (1,0) circle (3pt);
 \end{tikzpicture}
 \quad
 .
\end{equation}
Their description in terms of series or pseudo-differential operators will be given just by the multiplication of the associated series $G$ and $Y$. 
Even though we distinguish, at first, the action of these operators by applying them on different dummy
variables, in the end a single variable can be associated to the internal line, since the product of exponents
of the same \(p^2/\mu^2\) can be combined in the exponentiation by the sum of the variables. 
Take as an example the case of the operator $YGY$, associated to the third example in
equation~\eqref{propstrani}: by taking the Cauchy product of the series and by using the sum $x_{123}=x_1+x_2+x_3$
\begin{equation*}
 Y(a, \partial_{x_1})G(a, \partial_{x_2})Y(a, \partial_{x_3}) f(x_{123})= \sum_{n\geq 0} (\upsilon \gamma \upsilon)_n \frac{\partial^n_{x_{123}}}{n!} f(x_{123}) \,.
\end{equation*}
Furthermore, we will see in the next section that, since each of the factor satisfies the same kind of
renormalisation group equation, the product satisfies a renormalisation group equation which allows for
the easy evaluation of its higher orders in~\(L\), so that, {\em in fine}, solving our equations written
at a given order is no more complex than for preceding studies of \SDe.

\section{WSD equations and the renormalisation group }\label{Schwinger-Dyson and renormalization group equations}

Our strategy to solve the model is to use the Ward--\SDe\ to extract the anomalous dimensions of the $2$ and
$3$ point functions and then generate all their higher orders in~\(L\) by means of the renormalisation
group equations.  This is specially important for the propagator, since our equation~\eqref{SDe_K} does
not directly give the anomalous dimension \(\gamma\) as its coefficient of~\(L\), but a sum of the
two first coefficients of the propagator.  Higher-order terms in~\(L\) will likewise give 
sums of two terms. It is only because, at a given order in the coupling, the right-hand side of the equation
is polynomial in~\(L\) that we can have definite values for all these coefficients. A nonzero value
for the higher orders in~\(L\) of the inverse propagator~\(\Gamma_2\) would give an \(\exp(-L)\)
contribution, which would produce a constant term in~\(\Gamma_2\): such a constant is in contradiction
with the renormalisation condition of our massless theory. The fact that a constant term is allowed by our
equation can be easily understood, since it is an equation for the derivative of~\(\Gamma_2\).

Deducing the renormalisation group equations from the usual renormalisation procedure is no
longer the option it was in preceding works like~\cite{KrYe2006,BeSc08}, since the propagators are
deduced from their derivatives and do no longer correspond to the evaluation of a set of Feynman
diagrams. Howeveri, we will see that the \RGe\ can be seen as consequences of the \SDe, in a rather simple
way, without any need for special properties of the combinatorial solution of the \SDe.  In particular,
the $\beta$-function appears naturally as a combination of different anomalous
dimensions and an ``effective coupling" can be obtained as a combination of two- and three-point functions.

We will start from a rather naive observation: we consider a formal series $A(a,L)$ over the ring of
functions in the variable $a$, which is supposed to satisfy a renormalisation group like equation with
the anomalous dimension $\gamma_A$:
\begin{equation}\label{renormA}
\D A(a,L) = (\gamma_A (a) + \beta a\partial_a) A(a,L).
\end{equation}
We have the following lemma: let $\mathcal{P}(\D)$ be a polynomial with constant coefficients in the
variable~$\D$, then if $A(a,L)$ satisfies the preceding equation~\eqref{renormA} 
\begin{equation} \label{renormAm}
\D \circ \mathcal{P}(\D) A^m(a,L) = (m\gamma_A (a) + \beta a\partial_a) \mathcal{P}(\D) A^m(a,L).
\end{equation}
We see that $\mathcal{P}(\D) A^m(a,L) $ satisfies a \RGe\ with $m\gamma_A$ as anomalous dimension.  This happens simply because the coefficients of $\mathcal{P}(\D) $ do not depend on $a$ and the $\D$ commutes with $(\gamma_A (a) + \beta a\partial_a)$.
The lemma could be even generalised if we consider another formal series $B(a,L)$ that satisfies \eqref{renormA} with an anomalous dimension $\gamma_B(a)$ but with the same $\beta$-function. We have 
\begin{equation}
\D \circ \mathcal{P}(\D) A^m(a,L)B^n(a,L) = (m\gamma_A (a) + n\gamma_B(a) + \beta a\partial_a) \mathcal{P}(\D) A^m(a,L)B^n(a,L)
\end{equation}
and now the anomalous dimension is $m\gamma_A (a) + n\gamma_B(a)$.

We will now apply that lemma to the two series we have used to represent the $2$ and the $3$ point
functions, respectively $G(a,L)$ and $Y(a,L)$. Supposing that they satisfy renormalisation group
equations with anomalous dimensions $\gamma$ and $\upsilon$ up to some order in \(a\) and~\(L\), we will
show that the Ward--{\SDe} allow to extend them to a higher order.

In order to make the formulas less
cumbersome, we will write $G_x$ instead of $G(a,\partial_x)$, with the dependency on $a$ understood.
We can write the general Schwinger--Dyson equation as a series of primitive graph contributions:
\begin{equation}\label{CSD}
	C^\nu (a,L) = 2p^\nu \sum_n a^n \sum_{\textbf{p} \in \textbf{P}_n}
	\biggl( \prod_{i=1}^{I-1} G_{x_i} \biggr) \left(\D G_{x_I} -G_{x_I}\right)
	\Bigl( \prod_{j=I+1}^{I+V} Y_{x_j} \Bigr) 
	e^{-\omega_\textbf{p} L} F_{\textbf{p}}\left(\{x_i\},\mu^2\right)
\end{equation}
with the special link with a derivative of the propagator in the $I$-th position.
The terms on the right-hand side are monomials of pseudo-differential operators applied to
some characteristic integrals depending on the chosen model. These integrals are similar to those
usually found in the literature, but they are evaluated in fixed dimension and appear with arbitrary
decorations of the propagators and not the usual ones, integers eventually shifted by multiples of the
dimensional parameter. They are generically divergent for \(\omega_\textbf{p} =\sum x_i\) equal to 0,
where the evaluation finally takes place, but this divergence is a simple pole which is compensated by
taking at least one derivative with respect to~\(L\). This is coherent with the renormalization
approach, since the value of the left-hand part for \(L\) null can be fixed at will. 

Our aim is to express the derivative with respect to~\(L\) of equation~\eqref{CSD}. First of all, we can
get rid of the special r\^ole of~\(x_I\) by using equation~\eqref{xmGm} to rewrite
$\left( \D G_{x_I} -G_{x_I} \right) = G_{x_I} (x_I -1)$ and include the term \( (x_I - 1) \), as well as
a factor \(\omega_\textbf{p}\), to suppress the divergence, into \(F_\textbf{p}\) to define ${\tilde{F}}_{\textbf{p}}$. Furthermore let us denote
\begin{equation}
\mathcal{O}_\mathbf{x} := \prod_{i=1}^I G_{x_i}  \prod_{j=I+1}^{I+V} Y_{x_j}
\end{equation}
and rewrite \eqref{CSD} as 
\begin{equation}
\D C^\nu (a,L) = 2p^\nu \sum_{ n\geq 1} a^n \sum_{\textbf{p} \in \textbf{P}_n} \mathcal{O}_\mathbf{x} \, e^{\sum x_i L} {\tilde{F}}_{\textbf{p}}\left(\mathbf{x},\mu^2\right).
\end{equation}
or equivalently 
\begin{equation}\label{CSD_2}
	\D^2 G^{-1} + \D G^{-1} = \sum_{ n\geq 1} a^n \sum_{\textbf{p} \in \textbf{P}_n} \mathcal{O}_\mathbf{x}
	\, e^{\sum x_i L} {\tilde{F}}_{\textbf{p}}\left(\mathbf{x},\mu^2\right).
\end{equation}
If we now take $\D$ on both sides, we have
\begin{eqnarray}
\D^3  G^{-1} + \D^2 G^{-1} &=&
\sum_n a^n \sum_{\textbf{p} \in \textbf{P}_n}  \mathcal{O}_\mathbf{x} \left(\sum x_i\right) \,  e^{\sum x_i L} {\tilde{F}}_{\textbf{p}}\left(\mathbf{x},\mu^2\right) \\
&=&  \sum_n a^n \sum_{\textbf{p} \in \textbf{P}_n} \sum_k \mathcal{O}_{\mathbf{x} \setminus x_k} \D
\mathcal{O}_{x_k}  \, e^{\sum x_i L} {\tilde{F}}_{\textbf{p}}\left(\mathbf{x},\mu^2\right).
\end{eqnarray}
Here $\mathcal{O}_{x_k}$ denotes either $G$ or $Y$ according to \(k\) and since both of them satisfy a
renormalisation group equation with the same $\beta$ according to our recurrence hypothesis, then 
\begin{equation}
	\sum_k \mathcal{O}_{\mathbf{x}\setminus k} \D \mathcal{O}_k = \left(-\gamma + (3\gamma +2\upsilon) n + \beta a\partial_a\right)\mathcal{O}_\mathbf{x}
\end{equation}
since at a given order $n$ we have $I=3n-1$ and $V=2n$. Defining $\beta = 3\gamma +2\upsilon$ and using
that $a\partial_a a^n \mathcal{O}_\mathbf{x} = a^n \left( n + a\partial_a \right) \mathcal{O}_\mathbf{x}$, 
we have  
\begin{equation}
	\D^2 C^\nu(a,L) =  \left(-\gamma + \beta  a\partial_a \right)  \D C^\nu(a,L).
\end{equation}
We need to now recognise that \(C\) is as an inverse propagator, so that the minus sign in front of
\(\gamma\) in the preceding equation is just the case \(m=-1\) of equation~\eqref{renormAm}. A similar
computation for the three point function would give similarly, using that in this case, at order~\(n\),
we have that \(I = 3n\) and \(V = 2n+1\), 
\begin{equation}
   \D^2 Y(a,L) = \left( \upsilon + \beta a \partial_a \right) \D Y(a,L).
\end{equation}
It seems that we are missing the case where the Green functions are taken at \(L=0\), but since the
renormalisation conditions imply that the function at \(L=0\) are simply constants independent on~\(a\),
this part of the renormalisation group equations are used to define \(\gamma\) and~\(\upsilon\). It is
the very fact that the \SDe\ can only define the derivative with respect to~\(L\) of the Green
functions, due to the divergence in the constant part, which introduces the possibility of a breaking of
the scale invariance of the theory through the introduction of the anomalous dimensions \(\gamma\)
and~\(\upsilon\), which will produce the \(\beta\)-function.

We therefore have accomplished our aim of obtaining the renormalization group equations at an arbitrary order
in the coupling~\(a\) from their lower order versions.  This establishes them at all
orders through the recurrence principle.  This proof has two essential ingredients, first, the reduction of
all primitive contributions in the right-hand side of Ward--\SDe\ to depend only on the propagator and
the single-scale version of the vertex, secondly, the simple exponential dependence on the scale~\(L\). This latest
point is the trickiest to generalise to other theories, since this simple scale dependence is only true if
infrared rearrangement does not introduce spurious divergences.  This could be a problem in QCD where it is known
that infrared rearrangement must be combined with a substraction of induced infrared divergences, see
e.g.,~\cite{FaChHeVe2018}.

Higher-point functions, which cannot be
primitively divergent, do not entail the introduction of new renormalisation group functions but also satisfy
their own renormalisation group equation, with a linear combination of \(\gamma\) and \(\upsilon\) as anomalous
dimension.  The three-point function for generic kinematic configuration will also be given by convergent
contributions obtained through the difference with respect to a single-scale configuration as in
equation~\eqref{deformationvertex}.

The derivation in this section should allow to consider the \SDe\ as an alternative approach to the
renormalisation of this theory. The only regularisation we use is a simple consequence of our need to
invoke propagators with arbitrary exponents in order to be able to consider the logarithmic corrections
to the propagators and vertices, in the spirit of analytic regularisation~\cite{Sp69,Sp71,Speer_1976},
with the simplification that we only consider it for primitively divergent diagrams.  The resulting Mellin
transform has only a single pole in the neighborhood of 0 which is cancelled by the differentiation with respect
to \(L\).

\section{Consistency check: no vertex correction}\label{no_vertex}
Before delving in the computations for our model, let us indulge first in a case already studied by
one of the author in~\cite{Be10a,Be10}, where no vertex correction is needed. It will be useful to set
up notations and to verify that we can recover usual results.

We consider a complex field with an interaction Lagrangian density proportional to $\phi^3 +\phi^{*3}$.
In this case, we do not have vertex corrections at the one loop approximation and it is consistent to
only consider the Schwinger--Dyson equation for the propagator.
We start from the equation 
\begin{equation}
\begin{tikzpicture}
\draw (-0.5,0) -- (0.5,0);
\filldraw [gray] (0,0) circle (3pt);
\end{tikzpicture}
= 
\begin{tikzpicture}
\draw (-0.5,0) -- (0.5,0);
\end{tikzpicture}
\;
- \frac{1}{2}\;\;
\begin{tikzpicture}
\draw (-0.8,0) -- (-0.5,0);
\draw (0.5,0) -- (0.8,0);
\draw (0.5,0) arc (0:-180:0.5);
\draw (0.5,0) arc (0:180:0.5);
\filldraw [black] (0, 0.5) circle (2pt);
\filldraw [black] (0, -0.5) circle (2pt);
\end{tikzpicture}
\end{equation}
that in integral form reads
\begin{equation}
\Gamma_2(a,L) = p^2 - \frac{g^2}{2} \int_{\mathbb{R}^6} \frac{du}{(2\pi)^6} P(a, \log\, u^2/\mu^2) P(a, \log\, (p+u)^2/\mu^2).
\end{equation}
This enforces an equation for the $G^{-1}(a,L)$ series:
\begin{equation}\label{checkdivergent}
G^{-1}(a,L) = 1 - \frac{a}{2} G_x G_y \, e^{(x+y)L}  \frac{\Gamma(-1-x-y) \Gamma(2+x)\Gamma(2+y)}{\Gamma(1-x)\Gamma(1-y)\Gamma(4+x+y) } .
\end{equation}
This expression is manifestly divergent in the neighbourhood of the origin due to the pole of
$\Gamma(-1-x-y)$ for \(x+y=0\). This divergence could be compensated by a mere differentiation with
respect to~\(L\), but other terms can be added. In particular, we can follow~\cite{Be10a} and add a
second derivative with respect to~\(L\), while in~\cite{Be10}, a third derivative was used. To simplify
notations, $\partial_L$, the partial derivative with respect to~\(L\), will imply an evaluation at~$L=0$.
Applying  $\partial_L + \partial_L^2$ to equation~\eqref{checkdivergent} result in the multiplication of
the second hand by \((x+y)+(x+y)^2\) and we get
\begin{equation}\label{checkSD}
(\partial_L + \partial_L^2) G^{-1}(a,L) = - \frac{a}{2} G_x G_y  H(x,y)
\end{equation}
with $H(x,y)$ regular in the neighbourhood of the origin given by
\begin{equation}\label{checkH}
H(x,y)= \frac{\Gamma(1-x-y) \Gamma(2+x)\Gamma(2+y)}{\Gamma(1-x)\Gamma(1-y)\Gamma(4+x+y) }.
\end{equation}

In our new scheme, we consider an equation for the derivative~\(C^\nu\), with a divergent second
member, so that we still have to do a further differentiation with respect to~\(L\) to get a finite
equation.  The left hand side, using the property~\eqref{Cdevelopement} of~\(C^\nu\), will also be given
as~\( (\partial_L + \partial_L^2) G^{-1}\).
We therefore start from
\begin{equation}
\begin{tikzpicture}
\draw (-0.5,0) -- (0.5,0);
\filldraw [gray] (-0.1,-0.1) -- (-0.1,0.1) -- (0.1,0.1) -- (0.1,-0.1);
\end{tikzpicture}
^\nu
= 
\begin{tikzpicture}
\draw (-0.5,0) -- (0.5,0);
\end{tikzpicture}
^\nu
\;\;
- \frac{1}{2}\;\;
\begin{tikzpicture}
\draw (-0.8,0) -- (-0.5,0);
\draw (0.5,0) -- (0.8,0);
\draw (0.5,0) arc (0:-180:0.5);
\draw (0.5,0) arc (0:180:0.5);
\filldraw [black] (0, 0.5) circle (2pt);
\filldraw[black] (-0.1,-0.4) -- (0.1,-0.4) -- (0.1,-0.6) --(-0.1,-0.6)--cycle;
\end{tikzpicture}
\end{equation}
or in integral form 
\begin{equation}
C^\nu (a,L) = 2p^\nu -   \frac{g^2}{2} \int_{\mathbb{R}^6} \frac{du}{(2\pi)^6} \  P(a, \log\, u^2/\mu^2) K^\nu(a, \log\, (p+u)^2/\mu^2)  .
\end{equation}
In this particular one loop case, the result can be obtained simply by taking the
derivative~$\partial^\nu$ of the equation~\eqref{checkdivergent} to obtain
\begin{equation} \label{checkC}
	C^\nu (a,L) = 2p^\nu \left( 1 -a G_x G_y \, e^{(x+y)L}
	\frac{\Gamma(-x-y) \Gamma(2-x) \Gamma(2-y)}{\Gamma(1-x)\Gamma(1-y)\Gamma(4+x+y)} \right).
\end{equation}
We still have a divergence coming from the pole in~\(\Gamma(-x-y)\), which can be cancelled by a
differentiation with respect to~\(L\). Using equation~\eqref{Cdevelopement} to express \(C^\nu\), we have 
\begin{equation}
	\partial_L^2 G^{-1} + \partial_L G^{-1} = -\frac{a}{2} G_xG_y H(x,y) 
\end{equation}
with the same $H(x,y)$ defined in equation~\eqref{checkH}, so that we obtain the same equations for
the propagator as in the previous case.  In the case of all other graphs, simple primitive ones described in
equation~\eqref{primitives} or combinations coming from the corrections to the infrared rearrangement
introduced in equation~\eqref{prop_corr}, such a simple derivation is not possible and therefore we
present a direct computation in appendix~\ref{appendix1} which can be generalised to these other
cases.  The Mellin transform of the graph with a propagator \(p^\nu (p^2)^{y-2}\) is the same
function appearing in equation~\eqref{checkC}, but for a~\(\Gamma(2-y)\) in the denominator. However,
the expression of~\(K^\nu\) in terms of~\(G\) in equation~\eqref{refKnu} introduces a factor \(y-1\)
which gives back the preceding result.

To solve equation~\eqref{checkSD}, we must express the derivatives of the inverse propagator in terms of
\(\gamma\) with the help of equation~\eqref{Gamma2development} and use the renormalisation group equation
for $G(a,L)$, which can be spelled out as the following recursion for the coefficients
\begin{equation}
\gamma_{n+1} = \gamma(1 + 3 a\partial_a) \gamma_n 
\end{equation}
since $\beta= 3\gamma$ when there is no correction to the vertices.
This ultimately brings us to 
\begin{equation}\label{checkgamma}
\gamma = \gamma (1 - 3 a\partial_a ) \gamma + \frac{a}{2} G_x G_y  H(x,y).
\end{equation}
Using the first derivative of~\(H(x,y)\), we have
\begin{equation}
G_x G_y H(x,y) =  \left( \frac{1}{6} -  \frac{5}{18} \gamma + \ldots \right)  
\end{equation}
and the solution starts as
\begin{equation}
\gamma = \frac{1}{12} a - \frac{11}{432} a^2 + \ldots
\end{equation}

\section{Computations}
It is now time to apply all the machinery to the full $\phi^3$ theory and check that it can reproduce
known results. This will involve calculating corrections to the propagator and to the vertex.
\subsection{Propagator: the cat}
Let us consider the first non-trivial term in the full Schwinger--Dyson equation \eqref{SDe_K}:
\begin{equation} \label{cat}
\begin{tikzpicture}
\draw (-0.8,0) -- (-0.5,0);
\draw (0.5,0) -- (0.8,0);
\draw (-0.5,0) -- (-0.6,0.3)-- (-0.4,0.3);
\draw (0.5,0) -- (0.6,0.3) -- (0.4,0.3);
\fill [gray] (-0.5,0) -- (-0.6,0.3) -- (-0.4,0.3);  
\fill [gray] (0.5,0) -- (0.6,0.3) -- (0.4,0.3);  
\fill [white] (0.5,0) arc (0:180:0.5);
\draw (0.5,0) arc (0:-180:0.5);
\draw (0.5,0) arc (0:180:0.5);
\filldraw [white] (0.1,-0.4) -- (-0.1,-0.4) -- (-0.1,-0.6) -- (0.1,-0.6)--cycle;
\filldraw[black] (-0.1,-0.4) -- (0.1,-0.4) -- (0.1,-0.6) --(-0.1,-0.6)--cycle;
\filldraw [black] (0,0.5) circle (2pt);
\end{tikzpicture}
\, := \, 
\int_{\mathbb{R}^6} \frac{du}{(2\pi)^6}  \, \Gamma_3(-u) P(u) \Gamma_3(u) K^\nu(u+p)
\end{equation}
The integral is identical to the one encountered in the previous section and gives rise to the same Mellin
transform~\(H(x,y)\), given in equation~\eqref{checkH}. We therefore obtain a quite similar equation for
\(\gamma\) apart for two differences.  The second derivative of the propagator now involves a \(\beta\)
function which depends on the vertex corrections and the vertex corrections introduce further terms
proportional to~\(\log(u^2/\mu^2)\).
We get a contribution for the anomalous dimension 
\begin{equation}
- \gamma_2 + 2\gamma^2 -\gamma = - \frac{a}{2} \left( YGY \right)_x G_y H(x,y).
\end{equation}
where we see now the product of series along the line with variable $x$. It is important to notice
that we could not have moved the vertex corrections to the line with a derivative of the propagator,
since the trick of obtaining the part with \(\D G\) in~\(K^\nu\) simply by a multiplication by \(y\)
would no longer work.
It is interesting to notice that the natural separation of variables allows us to think of the
diagram as composed of the two following propagator-like objects: 
\begin{equation*}
 \begin{tikzpicture}
  \draw (0,0) -- (2,0);
  \filldraw [white] (0.8,-0.2) -- (1.2,-0.2) -- (1.2,0.2) -- (0.8, 0.2)--cycle;
  \filldraw[black] (0.8,-0.2) -- (1.2,-0.2) -- (1.2,0.2) -- (0.8, 0.2)--cycle;
 \end{tikzpicture}
 \quad
 \begin{tikzpicture}
  \draw (0,0) -- (2,0);
  \fill [gray] (0,0) -- (0.25,0.4) -- (0.5,0);  
  \fill [gray] (1.5,0) -- (1.75,0.4) -- (2,0);  
 \end{tikzpicture}
\end{equation*}

\subsection{Vertex: the rising sun}
We must also look at the vertex contribution. In equation~\eqref{vertex0}, we look for the lowest
order correction and use infrared rearrangement to obtain a one-loop diagram quite similar to the one
studied in the previous section, only simpler since it does not involve \(K^\nu\).
We look therefore at
\begin{equation}
\begin{tikzpicture}
    \fill [gray] (0.6,0) -- (0.45,-0.3) -- (0.3,0);      
	\fill [gray] (-0.6,0) -- (-0.45,-0.3) -- (-0.3,0); 
	\fill [gray] (0,0.6) -- (-0.2, 0.8) -- (-0.26, 0.55);
    \fill [white] (0.6,0) arc (0:180:0.6);
	\draw (-1,0) -- (1,0);
	\draw (0.6, 0) arc (0:180:0.6);
	\draw [dashed] (0,0.6) -- (0,1);
	\filldraw [black] (0, 0) circle (2pt);       
	\filldraw [black] (0.42, 0.42) circle (2pt);
	\filldraw [black] (-0.42, 0.42) circle (2pt);
\end{tikzpicture}
\, := \, 
\int_{\mathbb{R}^6} \frac{du}{(2\pi)^6}  \, P(u) \Gamma_3(u) P(u) \Gamma_3(u+p) P(u+p) \Gamma_3(u+p)
\end{equation}
where the dashed line describes a $0$-momentum coming in. We will see that the position of the vertex
corrections does not modify the value of this diagram at the approximation level we compute in this
work, but we argue that it is nevertheless the good one.  A first, quite mundane argument is that it
allows to reuse a propagator-like object already appearing in the previous section. The more serious
one is that in the asymptotic analysis we plan to do in the near future~\cite{BeRu20a}, this is the
form which allows for the clearer derivation.  This diagram is divergent by power counting, so we
expect a contribution to the anomalous dimension $\upsilon$.
Looking at the first coefficient in~\(L\) of equation~\eqref{vertex0}, we have 
\begin{equation}
\upsilon = - a \left( GYG \right)_x \left( YGY \right)_y \, H_2(x,y) .
\end{equation}
with the new Mellin transform~\(H_2(x,y)\) given by 
\begin{equation}
H_2(x,y) = \frac{\Gamma(1-x-y)\Gamma(1+x)\Gamma(2+y)}{\Gamma(2-x)\Gamma(1-y)\Gamma(3+x+y)}
\end{equation}
Again in this case we can imagine the diagram made of two composite objects:
\begin{equation*}
 \begin{tikzpicture}
  \draw (0,0) -- (2,0);
  \fill [gray] (0.75,0) -- (1,0.4) -- (1.25,0);  
 \end{tikzpicture}
 \quad
 \begin{tikzpicture}
  \draw (0,0) -- (2,0);
  \fill [gray] (0,0) -- (0.25,0.4) -- (0.5,0);  
  \fill [gray] (1.5,0) -- (1.75,0.4) -- (2,0);  
 \end{tikzpicture}
\end{equation*}
where we have omitted the dashed lines.

\subsection{First primitive diagram approximation}\label{First primitive diagram approximation}
We are now ready to set up a system of equation that will allow us to solve the theory in the
approximation of one primitive diagram for either of the propagator and vertex corrections. We have
\begin{equation}
 \begin{cases}
    \hspace{0.7cm} \mathcal{SD} : \hspace{2cm} &
	\begin{tikzpicture}
	\draw (-0.5,0) -- (0.5,0);
	\filldraw [gray] (-0.1,-0.1) -- (-0.1,0.1) -- (0.1,0.1) -- (0.1,-0.1);
	\end{tikzpicture}
	^\mu
	= 
	\begin{tikzpicture}
	\draw [black](-0.5,0) -- (0.5,0);
	\end{tikzpicture}
	^\mu
	- \frac{1}{2}\;\;
	\begin{tikzpicture}
	\draw (-0.8,0) -- (-0.5,0);
	\draw (0.5,0) -- (0.8,0);
	\draw (-0.5,0) -- (-0.6,0.3)-- (-0.4,0.3);
	\draw (0.5,0) -- (0.6,0.3) -- (0.4,0.3);
	\fill [gray] (-0.5,0) -- (-0.6,0.3) -- (-0.4,0.3);  
	\fill [gray] (0.5,0) -- (0.6,0.3) -- (0.4,0.3);  
	\fill [white] (0.5,0) arc (0:180:0.5);
	\draw (0.5,0) arc (0:-180:0.5);
	\draw (0.5,0) arc (0:180:0.5);
	\filldraw [black] (0, 0.5) circle (2pt);
	\filldraw [white] (-0.1,-0.4) -- (0.1,-0.4) -- (0.1,-0.6) --(-0.1,-0.6)--cycle;
	\filldraw[black] (-0.1,-0.4) -- (0.1,-0.4) -- (0.1,-0.6) --(-0.1,-0.6)--cycle;
	\end{tikzpicture} 
	\\
	\\
	\hspace{0.7cm}\mathcal{SD}  : &
	\begin{tikzpicture}
	\draw (-0.5,0) -- (0.5,0);
	\draw [dashed] (0,0) -- (0,0.5);
	\filldraw [gray] (0, 0) circle (3pt);
	\end{tikzpicture} 
	= 
	\begin{tikzpicture}
	\draw (-0.5,0) -- (0.5,0);
	\draw [dashed] (0,0) -- (0,0.5);
	\end{tikzpicture}
	+ 
    \begin{tikzpicture}
    \fill [gray] (0.6,0) -- (0.45,-0.3) -- (0.3,0);      
	\fill [gray] (-0.6,0) -- (-0.45,-0.3) -- (-0.3,0); 
	\fill [gray] (0,0.6) -- (-0.2, 0.8) -- (-0.26, 0.55);
    \fill [white] (0.6,0) arc (0:180:0.6);
	\draw (-1,0) -- (1,0);
	\draw (0.6, 0) arc (0:180:0.6);
	\draw [dashed] (0,0.6) -- (0,1);
	\filldraw [black] (0, 0) circle (2pt);       
	\filldraw [black] (0.42, 0.42) circle (2pt);
	\filldraw [black] (-0.42, 0.42) circle (2pt);
	\end{tikzpicture}
	\\
	\\
	\hspace{0.7cm}\mathcal{RG}  :  &\gamma_{n+1} = \left(\gamma_1 + \beta a\partial_a \right) \gamma_n 
	\\
	\mathcal{SD/RG}  : & \beta = 3\gamma + 2\upsilon
 \end{cases}
\end{equation}
Using the results of the preceding subsections, this can be converted in the following system of
equations:
\begin{equation}\label{systemtheory}
\begin{cases}
 2\gamma^2- \gamma_2 -\gamma = - \frac{a}{2} \left( YGY \right)_x G_y \, H(x,y) \\
\upsilon = -a \left( GYG \right)_x \left( YGY \right)_y \, H_2(x,y) \\
 \gamma_2 = (\gamma + \beta a\partial_a) \gamma \\
 \beta=3\gamma + 2\upsilon.
\end{cases}
\end{equation}
We do not write explicitly the equations needed to express the higher orders of~\(G\) or~\(Y\), since
they would only appear in computations at higher order, which could only be made exact through the
introduction of many more primitives than the one we will consider here.  At this stage, we simply
indicate that it would be much more efficient to directly compute the products \(GYG\) and~\(YGY\)
from appropriate version of the renormalisation group equations than to compute explicitly the
products.
The first step is to expand the right hand sides in the system~\eqref{systemtheory}.  We limit
ourselves to the linear terms in the functions \(H\) and~\(H_2\):
\begin{eqnarray}
	\left( YGY \right)_x \left( G \right)_y \, H(x,y)
	&=& \left( 1+(2\upsilon+\gamma)\partial_x + \dots \right)
		\left( 1+\gamma \partial_y+\dots \right) H(x,y)  \nonumber \\
	&=&\frac{1}{6}-\frac{5}{18}(\upsilon+\gamma) + \ldots ;  \\
\left( GYG \right)_x \left( YGY \right)_y  \,H_2(x,y)
	&=& \left( 1+(\upsilon+2\gamma)\partial_x +\dots \right)
	\left( 1+(\gamma+2\upsilon) \partial_y +\dots \right) H_2(x,y)  \nonumber \\
	&=& \frac{1}{2} -\frac{3}{4}(\upsilon+\gamma) +\ldots  .
\end{eqnarray}
We can now write the equations in~\eqref{systemtheory} as 
\begin{equation}
\begin{dcases}
\gamma =  \left( \gamma  -  \beta a\partial_a \right) \gamma + \frac{a}{12}\left(1-\frac{5}{3}(\upsilon+\gamma) + \ldots \right) \\
\upsilon = - \frac{a}{2} \left(1-\frac{3}{2}(\upsilon+\gamma) +\ldots \right) \\ 
\beta=3\gamma + 2\upsilon.
\end{dcases}
\end{equation}
The first two steps of the solution of this system are
\begin{align}\label{prelim}
1st: \hspace{0.8cm} & \gamma = \frac{a}{12} ; \quad \upsilon=-\frac{a}{2};\quad \beta=-\frac{3}{4}a\\
2nd : \hspace{0.8cm} & \gamma = \frac{a}{12}+\frac{55}{432} a^2 ; \quad \upsilon=-\frac{a}{2}-\frac{5}{16}a^2;\quad \beta=-\frac{3}{4}a-\frac{35}{144}a^2
\end{align}

\section{Corrections to the order $a^2$}\label{IRcorrections}
The results from the previous paragraph are not complete at order $a^2$. We have two corrections to consider: one to the anomalous dimension $\gamma$ and one to the anomalous dimension $\upsilon$.
\subsection{From the propagator}\label{fromthepropagator}
We first have the second term in \eqref{prop_corr}, which is of minimum order 2:
\begin{equation}\label{correctionstopropagator}
\begin{tikzpicture}[scale=1.2]
\draw (0.8,0) -- (1,0);
\draw (-1,0) -- (-0.8,0);
\fill [white] (-0.3,0.5) arc (90:270:0.5);
\draw (-0.3,0.5) arc (90:270:0.5);
\fill [gray] (-0.8,0) -- (-0.9,0.3) -- (-0.7,0.3);  
\draw (0.3,0.5) arc (90:-90:0.5);
\draw (-0.3,0.5) -- (0.3,0.5);
\draw (-0.3,-0.5) -- (0.3,-0.5);
\filldraw [white] (-0.3,-0.4) -- (-0.1,-0.4) -- (-0.1,-0.6) --(-0.3,-0.6)--cycle;
\filldraw[black] (-0.3,-0.4) -- (-0.1,-0.4) -- (-0.1,-0.6) --(-0.3,-0.6)--cycle;
\draw (0.3,0.5) -- (0.3,-0.5); 
\fill [gray] (0.8,0) -- (0.9,0.3) -- (0.7,0.3);  
\fill [gray] (0.3,0.5) -- (0.3,0.2) -- (0.15,0.35);  
\fill [gray] (0.3,-0.5) -- (0.3,-0.2) -- (0.15,-0.35);  
\end{tikzpicture}
\quad
-
\quad
\begin{tikzpicture}[scale=1.2]
\draw (-1,0) -- (-0.8,0);
\fill [white] (-0.3,0.5) arc (90:270:0.5);
\draw (-0.3,0.5) arc (90:270:0.5);
\filldraw [gray] (0.8,0.2) -- (0.8,-0.1) -- (1,0.05) -- (0.8,0.2); 
\fill [white] (0.4,0.5) arc (90:-90:0.45);
\draw (0.4,0.5) arc (90:-90:0.45);
\draw (-0.3,0.5) -- (0.4,0.5);
\draw (-0.3,-0.5) -- (0.3,-0.5);
\filldraw [white] (-0.3,-0.4) -- (-0.1,-0.4) -- (-0.1,-0.6) --(-0.3,-0.6)--cycle;
\filldraw[black] (-0.3,-0.4) -- (-0.1,-0.4) -- (-0.1,-0.6) --(-0.3,-0.6)--cycle;
\fill [gray] (-0.8,0) -- (-0.9,0.3) -- (-0.7,0.3);  
\draw (0.4,-0.4) -- (1,-0.4);
\draw (0.4,-0.4) arc (0:-90:0.1);
\draw (0.4,0.5) arc (90:-90:0.45);
\draw (0.4,-0.45) -- (0.4,0.5);
\fill [gray] (0.4,0.5) -- (0.4,0.2) -- (0.25,0.35);  
\fill [gray] (0.4,-0.45) -- (0.4,-0.15) -- (0.25,-0.35);  
\end{tikzpicture}
\quad
\sim
a^2
\end{equation}
To estimate this contribution let us close the diagrams by connecting the external lines with such a
propagator that they
become conformal \footnote{In a forthcoming paper, there will be a more complete characterisation of the method~\cite{Be20}} :
\begin{equation}
\begin{tikzpicture}
\filldraw [gray] (-1,0) -- (-1.2,.4) -- (-.8,.4) --cycle;
\filldraw [gray] (1,0) -- (1.2,.4) -- (.8,.4) --cycle;
\filldraw[white] (0,0) circle (1);
\filldraw[gray] (0,1) -- (-.26,.85) -- (0,.7) --cycle;
\filldraw[gray] (0,-.5) -- (-.26,-.35) -- (0,-.2) --cycle;
\draw (0,0) circle (1);
\draw (-1,0) arc (216.87:323.13:1.25);
\filldraw[black] (-.6,-.3) -- (-.4,-.3) -- (-.4,-.5) -- (-.6,-.5)--cycle;
\draw (0,1) -- (0,-.5);
\end{tikzpicture}
\quad
-
\quad
\begin{tikzpicture}
\filldraw [gray] (-1,0) -- (-1.2,.4) -- (-.8,.4) --cycle;
\filldraw [gray] (.8,.3) -- (1,.8) -- (.6,.8) --cycle;
\filldraw[white] (0,0) circle (1);
\draw (0,0) circle (1);
\draw (-1,0) arc (216.87:323.13:1.25);
\filldraw[black] (-.1,-.4) -- (.1,-.4) -- (.1,-.6) -- (-.1,-.6)--cycle;
\draw (1,0) -- (-.5,.866);
\filldraw[gray] (-.5,.866) -- (-.55, .55) -- (-.25,.72)--cycle;
\filldraw[gray] (.73,.16) -- (.72, -.14) -- (1,0);
\end{tikzpicture}
\end{equation}
The propagator with the square is of the form $p^\mu/(p^2)^2$ so the line we added will carry a
similar factor~\(q^\mu\) to make the diagram a scalar. To fix notations, the lines 1, 2 and~3 form the
triangular subdivergence, 5 is the line carrying the box decoration and 6 is the line added to close
the graph.
We call $\Gamma_{\,\widetilde i}$ the completed graphs and $\Gamma_i$ the propagator-like ones.
Massless vacuum diagrams have a vanishing second Symanzik polynomial and the first one is related to
the Symanzik polynomials of the uncompleted graph. Calling the Schwinger parameters $t_i$, we have
\begin{equation}
\begin{tikzpicture}
\filldraw [gray] (-1,0) -- (-1.2,.4) -- (-.8,.4) --cycle;
\filldraw [gray] (1,0) -- (1.2,.4) -- (.8,.4) --cycle;
\filldraw[white] (0,0) circle (1);
\filldraw[gray] (0,1) -- (-.26,.85) -- (0,.7) --cycle;
\filldraw[gray] (0,-.5) -- (-.26,-.35) -- (0,-.2) --cycle;
\draw (0,0) circle (1);
\draw (-1,0) arc (216.87:323.13:1.25);
\filldraw[black] (-.6,-.3) -- (-.4,-.3) -- (-.4,-.5) -- (-.6,-.5)--cycle;
\draw (0,1) -- (0,-.5);
\end{tikzpicture}
\quad \quad
\psi_{\, \widetilde{1}} = t_6 \psi_{\, 1} + \phi_{\, 1}.
\end{equation}
and
\begin{equation}
\begin{tikzpicture}
\filldraw [gray] (-1,0) -- (-1.2,.4) -- (-.8,.4) --cycle;
\filldraw [gray] (.8,.3) -- (1,.8) -- (.6,.8) --cycle;
\filldraw[white] (0,0) circle (1);
\draw (0,0) circle (1);
\draw (-1,0) arc (216.87:323.13:1.25);
\filldraw[black] (-.1,-.4) -- (.1,-.4) -- (.1,-.6) -- (-.1,-.6)--cycle;
\draw (1,0) -- (-.5,.866);
\filldraw[gray] (-.5,.866) -- (-.55, .55) -- (-.25,.72)--cycle;
\filldraw[gray] (.73,.16) -- (.72, -.14) -- (1,0);
\end{tikzpicture}
\quad \quad
\psi_{\, \widetilde{2}} = t_6 \psi_{\, 2} + \phi_{\, 2}.
\end{equation}
We point out that $\psi_{\, 1}=\psi_{\, 2}$, thus 
\begin{equation}\label{psidiff}
\psi_{\, \widetilde{1}} -\psi_{\, \widetilde{2}} = \phi_{\, 1} - \phi_{\, 2}.
\end{equation}
so in particular it does not depend on $t_6$.
In the absence of any decoration, for a zero momentum insertion on the line 5, we would calculate
\begin{equation}
\int dt_1 \ldots dt_6  t_5 t_6^{2}
\left( \frac{1}{\psi_{\, \widetilde{1}}^3} - \frac{1}{\psi_{\,\widetilde{2}}^3}\right)\delta_H
\end{equation}
while with the numerators we have
\begin{equation}\label{corrections_a2}
\int dt_1 \ldots dt_6 t_5  t_6^{3} \left( \frac{  C_{\, \widetilde{1}} }{{  \psi^4_{\, \widetilde{1} }} }
- \frac{  C_{\, \widetilde{2}} } { { \psi^4_{\, \widetilde{2} } }}\right) \delta_H.
\end{equation}
The \(\delta_H\) factors in these integrals represent the restriction of these integrations to a
hyperplane necessary to make these scale invariant integrals finite.  We will not further precise these factors since these
homogeneous integrals are independent on their precise choice.
The exact formulas involve \(\Gamma\) factors which become important when considering modified
propagators, but at this stage we can ignore them, because they are either taken for an index 1 or 2
and give 1, or there is a compensation between a \(\Gamma(D/2)\) factor in the numerator and the same
factor in the denominator coming from the exponent of \(t_6\).
The computation of the numerators was completely described in a book by Nakanishi~\cite{Na71}:
they could be calculated as a minor of a matrix associated to the graph, or by finding subgraphs
with exactly one cycle including the lines 5 and~6.  We have
\begin{align}
C_{\, \widetilde{1} } &= t_4 (t_1 + t_2 + t_3) + t_3 t_{2} \\
C_{\, \widetilde{2} } &= t_4 (t_1 + t_2 + t_3) + (t_1 + t_2)t_3
\end{align}
and therefore
\begin{equation}\label{c1_c2}
C_{\, \widetilde{1} }-C_{\, \widetilde{2} }= -t_1t_3.
\end{equation}
We can therefore write the parenthesis in equation~\eqref{corrections_a2} as:
\begin{align}
\frac{C_{\, \widetilde{1}}}{ {  \psi^4_{\, \widetilde{1} } } } - \frac{     C_{\, \widetilde{2} } } { {   \psi^4_{\, \widetilde{2} } }}&= 
 C_{\, \widetilde{1}} \left( \frac{1}{\psi^4_{\, \widetilde{1} }} - \frac{1}{\psi^4_{\, \widetilde{2} }}
 \right) -  \frac{t_1t_3}{\psi^4_{\, \widetilde{2}}}  \nonumber \\ \label{corr_integrand}
&= C_{\, \widetilde{1} } \left( \psi_{\, \widetilde{2}}- \psi_{\, \widetilde{1}}   \right) 
\left( \frac{1}{\psi^4_{\, \widetilde{1}}\psi_{\, \widetilde{2}}} + \frac{1}{\psi_{\, \widetilde{1}}\psi^4_{\, \widetilde{2}}} + \frac{1}{\psi^2_{\, \widetilde{1}}\psi^3_{\, \widetilde{2}}} + \frac{1}{\psi^3_{\, \widetilde{1}}\psi^2_{\, \widetilde{2}}}  \right) - \frac{t_1t_3}{\psi^4_{\, \widetilde{2}}}.
\end{align}
From equation~\eqref{psidiff} we know that $\left( \psi_{\, \widetilde{2}}- \psi_{\, \widetilde{1}}
\right) $ does not depend on $t_6$ so that the big first term in the preceding equation behaves as
\(1/t^5_6\) for large \(t_6\),  making therefore a convergent contribution in
equation~\eqref{corrections_a2} which will be absorbed in the renormalisation
condition.\footnote{Since the difference we look at makes a primitive divergence, this global
divergence is the only one which has to be checked, but one can also explicitly see that there are
no subdivergences associated to the $(1,2,3)$ subdiagram.}
At order~\(a^2\), we are left with the contribution coming from $-t_1t_3t_5t_6^3/\psi^4_{\widetilde{2}}$.
This integral is the one we would obtain from the graph $\Gamma_{\tilde{2}}$ in $D=8$ since the
exponent of the denominator is $4$. The factors in the numerator imply, however, that it must be taken 
with propagators with different indices.
\begin{equation}
\begin{tikzpicture}
\draw (1,0) arc (0:90:1) node[midway, right] {1};
\draw (0,1) arc (90:180:1) node[near start, above ] {1} node [near end, left] {2};
\draw (0,1) arc (0:-90:1) node[midway,right] {2}; 
\draw (1,0) arc (0:-180:1) node[midway,below] {4};
\draw (-1,0) arc (216.87:323.13:1.25) node[midway,above] {2};
\end{tikzpicture}
\end{equation}
We can proceed by reduction as shown in the appendix $\ref{appendix2} $
\begin{equation}
\begin{tikzpicture}
\draw (1,0) arc (0:90:1) node[midway, right] {1};
\draw (0,1) arc (90:180:1) node[near start, above ] {1} node [near end, left] {2};
\draw (0,1) arc (0:-90:1) node[midway,right] {2}; 
\draw (1,0) arc (0:-180:1) node[midway,below] {4};
\draw (-1,0) arc (216.87:323.13:1.25) node[midway,above] {2};
\end{tikzpicture}
\approx
\begin{tikzpicture}
\draw (1,0) arc (0:90:1) node[midway, right] {1};
\draw (0,1) arc (90:180:1) node[midway,left] {3};
\draw (0,1) arc (0:-90:1) node[midway,right] {2}; 
\draw (1,0) arc (0:-180:1) node[midway,below] {4};
\draw (-1,0) arc (216.87:323.13:1.25) node[midway,above] {2};
\end{tikzpicture}
\approx
\frac{\Gamma(0)}{\Gamma(4)} 
\quad
\begin{tikzpicture}
\draw (1,0) arc (0:360:1) node[near start, below] {3} node[near end, above] {3} ;
\draw (-1,0) -- (1,0) node[midway, above]{2};
\end{tikzpicture}
\approx 
\frac{\Gamma(0)}{24} 
\quad
\begin{tikzpicture}
\draw (.5,0) arc (0:360:.5) node[near start, below] {3} node[near end, above] {1};
\end{tikzpicture}
\end{equation}
We consider the last graph as the natural by-product of closing the diagram and therefore evaluate it to 1. 
The factor \(\Gamma(0)\) represents the divergence which will be compensated by taking the
derivative with respect to~\(L\), so that we remain with a residue $1/24$ which is the period of the
graph.  Finally, the \(1/2\) symmetry factor and the doubling associated to the presence of the two
symmetric possibilities for the corrected vertex cancel, so that we end up with a contribution \(-1/24
a^2\) for \(\gamma\).
\subsection{From the vertex}
In the non-trivial term appearing in equation~\eqref{vertex0}, the $4$-point function can be expanded as
in equation~\eqref{4ptfunction} to give
\begin{equation}
\begin{tikzpicture}
\draw[dotted, thick] (-0.8,0) -- (-0.5,0);
\draw  (-0.5,0) -- (0,0.5) -- (0.8, 0.5);
\draw (-0.5,0) -- (0,-0.5)-- (0.8, -0.5);
\filldraw [gray] (-0.5,0) circle (3pt);
\fill[gray] (0.4,0) ellipse (0.2 and 0.6);
\end{tikzpicture}
= 
\begin{tikzpicture}
\draw[dotted, thick] (-0.8,0) -- (-0.5,0);
\draw  (-0.5,0) -- (0,0.5) -- (0.8, 0.5);
\draw (-0.5,0) -- (0,-0.5)-- (0.8, -0.5);
\filldraw [gray] (-0.5,0) circle (3pt);
\draw (.4,.5)--(.4,-.5);
\filldraw [gray] (.4,.5) circle (3pt);
\filldraw [gray] (.4,-.5) circle (3pt);
\end{tikzpicture}
+
\frac{1}{2}
\quad
\begin{tikzpicture}
\draw[dotted, thick] (-0.8,0) -- (-0.5,0);
\draw  (-0.5,0) -- (0,0.5) -- (0.8, 0.5);
\draw (-0.5,0) -- (0,-0.5)-- (0.8, -0.5);
\filldraw [gray] (-0.5,0) circle (3pt);
\draw (.2,.5)--(.6,-.5);
\filldraw [white] (.3,0) circle (4pt);
\draw (.2,-.5) -- (.6,.5);
\filldraw [gray] (.2,.5) circle (3pt);
\filldraw [gray] (.6,-.5) circle (3pt);
\filldraw [gray] (.2,-.5) circle (3pt);
\filldraw [gray] (.6,.5) circle (3pt);
\end{tikzpicture}
\ldots
\end{equation}
A correction comes from the second term, in which, according to our approximation scheme, we can shift the
decorations of the vertices to obtain 
\begin{equation}\label{correctionstovertex}
\frac{1}{2}
\quad
\begin{tikzpicture}
\filldraw [gray] (0,1) -- (.3,.8) -- (.15, 1.2) -- cycle;
\filldraw [white]  (1,0) arc (0:180:1);
\filldraw [gray] (1,0) -- (.8,-.15) -- (.7,.09) -- cycle ; 
\filldraw [gray] (-1,0) -- (-.8,-.15) -- (-.7,.09) -- cycle ; 
\filldraw [gray] (-.8,.6) -- (-.7,.35) -- (-.5,.5) -- cycle ; 
\filldraw [gray] (.8,.6) -- (.7,.35) -- (.5,.5) -- cycle ; 
\draw (-1.2,0) -- (-1,0);
\draw (1.2,0) -- (1,0);
\draw (1,0) arc (0:180:1);
\draw (-.8,.6) -- (1,0);
\filldraw[white] (0,.3) circle (4pt);
\draw (-1,0) -- (.8,.6);
\draw[dotted, thick] (0,1.2) -- (0,1);
\end{tikzpicture}
\end{equation}
In fact, at the order $a^2$ we can simply ignore the decorations. 
We now close the diagram to obtain a vacuum one without any divergent subgraph
\begin{equation}
\frac{1}{2}
\quad
\begin{tikzpicture}
\draw (1,0) arc (0:360:1);
\filldraw [draw=black, fill=white] (0,.7) circle (.8);
\draw (0,1.5) -- (0,.8) node [right] {2} --(0,-.1);
\draw (0,-1) node  [above] {3};
\end{tikzpicture}
\end{equation}
so that the divergence will come only from the integration over $t_6$. In fact
\begin{equation}
\frac{1}{2} \int_{\mathbb{R}^6_+} dt_1\ldots dt_6 \frac{ t_5 t_6^2} {\tilde{\psi}^3} \delta_H
\end{equation}
where again 
\begin{equation}
\tilde{\psi} = t_6 \psi + \phi
\end{equation}
The divergence takes the form of a pole with a residue linked to the factorisation of the diagram in the
additional line 6 and the original diagram without its exterior lines.  In the context of fixed dimension
analytic regularisation, such poles have been studied in the general case in~\cite{BeSc12} and the residue
is obtained as \(1/\Gamma(D/2)\) times the product of the evaluation of the parts.  The single loop as
well as the remaining two loop graphs with all propagators with index 2 evaluate to 1, so that this
residue is just \(1/\Gamma(D/2)= 1/2\). 
With the symmetry factor \(1/2\) of this graph, we have therefore a contribution of $a^2/4$ for the anomalous
dimension~$\upsilon$.

At the same order, we seem to have an analog of the contribution studied in the preceding subsection, coming
from a correction to the infrared rearrangement.  However, this contribution turns out to be finite, like
the term proportional to \(C_1\) in equation~\eqref{corr_integrand}, so that it does not affect the anomalous
dimension~\(\upsilon\).

\subsection{Comparison with known results}
Our results are easily generalisable to include multicomponent fields. The structure of the diagrams 
remains unchanged and likewise the structure of the Schwinger--Dyson equations. The dependence on a single
coupling can be obtained if we have a symmetry group.  We would need to include
Casimir factors that in the literature are usually denoted by $T_i$. Starting from an interaction term
where $\phi^3$ is replaced by $d_{ijk} \phi^i \phi^j \phi^k$, the $T_i$ relate the trace of the $i$-fold
product of tensors $d_{ijk}$ and \(\delta_{ij}\) (in the case of \(T_2\)) or \(d_{ijk}\) (in the case of all
others \(T_j\), where \(j\) is always odd). 
The first two Casimir \(T_2\) and~\(T_3\) are defined by
\begin{equation}
d_{ijk} d_{ljk} = T_2 \delta_{il} \qquad d_{ilm} d_{jln} d_{kmn} = T_3 d_{ijk}.
\end{equation}
In fact, the order is not sufficient to characterize these Casimir factors, starting at order 7, but since
they would only appear in vertex corrections with at least three loops or propagator corrections with at
least one additional loop, this is of no concern to us in this work.
These group factors change our equations in the following way
\begin{align}
\gamma &= \left( \gamma -\beta a \partial_a \right) \gamma + \frac{T_2 a}{2} \left( YGY \right)_x G_y H(x,y) \\
\upsilon &= - T_3 a \left( GYG \right)_x \left( YGY \right)_z H_2 (x,z) \\
\beta &= 3 \gamma +2 \upsilon 
\end{align}
Since the operators contain themselves products of $\gamma$ and $\upsilon$ we expect products of $T_i$ appearing all over. 
Our first correction from \eqref{correctionstopropagator} will intervene with a factor $T_2T_3$ while the second one from \eqref{correctionstovertex} will come with a factor $T_5$.
With these corrections we have
\begin{align}
\gamma &= \frac{T_2}{12} a + \left(  -11 T_2   +  48 T_3     \right)  \frac{T_2}{432} a^2 \\
\upsilon &= - \frac{T_3}{2} a + \left( -\frac{T_5}{4}+ \frac{T_3}{16} \left( T_2 - 6T_3  \right) \right) a^2 \\
\beta &= \left( T_2 - 4 T_3 \right)  \frac{a}{4}  + \left( -11 T_2^2 + 66 T_2T_3 -108 T_3^2 -72 T_5 \right) \frac{a^2}{144}
\end{align}

We can compare it with previous results for the $\beta$-function~\cite{Gra2015,PaRoSu2016}. This
comparison is not immediate, since our definition of the $\beta$-function is unusual, as can be seen from
the way we have written the renormalisation group equations: $\beta (a) := \mu \frac{d}{d\mu} \log a=
\frac{2}{g}\mu \frac{d}{d\mu}g$. The coefficient of \(a\) is twice the coefficient of \(g^3\) in other
works and the one of \(a^2\) is twice the coefficient of~\(g^5\).  Furthermore, in the
work~\cite{Gra2015}, the conventions are such that the terms with odd powers of \(a\) have an additional
minus sign. With these provisions, we recover the usual \(\beta\)-function, but the \(T_2T_3\) term
in~\(\gamma\) is off by a factor of 2: this is not completely unexpected, since our renormalisation
conditions are different and only the \(\beta\)-function is scheme independent at this order. 

Further terms in these expansions could be interesting, but their comparison with published results would not be
straightforward.  Indeed, most computations are carried out in dimensional regularisation and
the information for their reduction to our renormalisation scheme is usually not provided even if it is necessary to make predictions on physical processes: we would need the finite
parts in the propagators and the vertex to relate the coupling constants in the two schemes.  Trying to carry our
computations in general dimensions in order to mimic dimensional regularisation would be much more demanding in
our scheme.

Nevertheless, reaching order four could be done in simple steps:  for the one-loop graphs, expanding the Mellin
transform to higher order; as for the graphs starting to contribute at these orders, they can be easily evaluated, using for
example
\texttt{HyperInt}~\cite{Pa2015}; for the already considered correction graphs, new derivatives of the Mellin
transform are needed, but most of them can be reduced to the tetrahedron graph in 6
and~8 dimensions which have a rich set of symmetries~\cite{Br86}.  The problem is much more about putting
together all components than evaluating any of the graphs.
In any case, this approach has the benefit of reducing the number of graphs that need to be considered since only primitive ones get included. 

\section*{Conclusions}
In this paper, we have established and solved the Ward--\SDe\ for a massless $\phi^3$ in $6$ dimensions.
Our interest in this model is motivated by the presence of vertex corrections, in the simple
context of a scalar theory, allowing to go beyond the Wess--Zumino model studied in our previous works.
The presence of overlapping divergences was a major block for the expression of
\SDe\ solvable in terms of renormalised Green functions. We have deployed a method inspired
by Ward~\cite{Wa51} that consisted in studying a version of the \SDe\  for the derivative
of the $2$-point functions with respect to the external momentum.  The reduction to analytically simple
versions of these equations was achieved by a deformation of the diagrams akin to the infrared
rearrangements used in other contexts (section §\ref{IR_rearrangement}).

An interesting part of this work is that in section~\ref{Schwinger-Dyson and renormalization group
equations}, we have shown that the \SDe\ imply the \RGe, bypassing the need to deduce them from the usual
renormalisation procedure. We could not develop it in this letter, but this could be the base of a
proof of the renormalized perturbative series independent on the methods of BPHZ~\cite{BoPa57,He66,Zi69}
and their more recent avatars~\cite{CoKr99,CoKr00}.  As in the works of Epstein
and Glaser~\cite{EpGl73}, fully renormalized lower order results are used to produce the next step
in the computation. The situation may become more complex in cases where infrared rearrangement leads to infrared divergences, luckily not the case for scalar theories. Infrared regulators could spoil the scale invariance and the dependence on $L$ of the propagator in equation~\eqref{CSD} might not be the simple exponential we had. 
It is one of the points which make the extension of this approach to gauge theories not straightforward.

Finally, in sections §\ref{First primitive diagram approximation} and~§\ref{IRcorrections} we have given the
solution up to the order $a^2$, fully compatible with known results.
It would be possible to proceed to higher orders, but we think that the present computations are
sufficient to illustrate the general procedure.  

We believe these methods could inspire applications to other models. In particular, we look forward to apply them to gauge theories, even if we are aware that it should not be straightforward.

\section*{Acknowledgement}
E.~Russo would like to thank S.~Teber for useful discussions about calculations of Feynman integrals.

\appendix
\section{Parametric representations}  \label{appendix1}
It is well known that it is possible to rewrite a Feynman integral as a projective integral.
The first step is to use the so called ``Schwinger trick":
\begin{equation}\label{Schwtrick}
x^{-\alpha}\Gamma(\alpha) = \int_\mathbb{R_+} dt \, t^{\alpha-1} e^{-tx}
\end{equation}
on each propagator.

Then we need some definitions. A graph is a collection of vertices and edges, a tree ($T$) is a graph with
no loops and a forest is a disjoint union of trees. A spanning tree (forest) is a tree (forest) which contains
all the vertices of the graph. Let $\mathcal{T}$ be the set of spanning trees and $\mathcal{F}_k$ the set
of spanning forests with exactly \(k\) components. Let $I$ be the set of internal edges $e_i$, $V$ the set
of vertices. We associate to each edge a mass parameter $m_i$ and a power of the propagator $\alpha_i$.
Finally let $p_T$ be the sum of the external momenta entering the subgraph \(T\). In the case of a graph
with only one connected component, $h_1(\Gamma) = |I| - |V| + 1$ is the number of loops.

It is then possible to associate to a given graph $\Gamma$ the two Symanzik polynomials
\(\psi_\Gamma\) and~\(\phi_\Gamma\) with the following definitions:
\begin{equation*}
\psi_{\Gamma}(\{t_i\}) := \sum_{T\in\mathcal{T}}\prod_{e_i\notin T}t_i
\end{equation*}
\begin{equation*} 
\phi_{\Gamma}\left(\{t_i\},\{p_i \cdot p_j\},\{m_i\}\right) :=
\sum_{(T_1,T_2)\in\mathcal{F}^2} p_{T_1}^2 
\biggl(\prod_{e_i\notin (T_1,T_2)}t_i\biggr) 
+ \psi_{\Gamma}(\{t_i\})\sum_{i=1}^{|I|}t_i{m^2_i}.
\end{equation*}
The Feynman integral for the graph \(\Gamma\) can then be written as
\begin{equation*} 
\mathscr{I}\left(\{p_i \cdot p_j\},\{\alpha_i\}, \{m_i\}, D\right) =  
  \int_{\mathbb{R}_+^{|I|}} \prod_{i\in I} \Bigl(\frac{dt_i t_i^{\alpha_i-1}}{\Gamma(\alpha_i)}\Bigr)  \frac{e^{-\phi /\psi}}{\psi^{D/2}}
\end{equation*}
where we have omitted the variables on which \(\psi\) and \(\phi\) depend for the sake of readability.
From their definition, we can see that the Symanzik polynomials \(\psi\) and~\(\phi\) are homogeneous
in the $t_i$ variables with respective degrees $h_1(\Gamma)$ and~$h_1(\Gamma) +1$. 
If we insert the equality
\begin{equation}\label{deltahyperplane}
1 = \int_0^{+\infty} d\lambda \, \delta(\lambda - H(t))
\end{equation}
for any nonzero hyperplane equation $H(t)= H^j t_j$, we obtain:
\begin{equation*}
\mathscr{I}\left(\{p_i\},\{\alpha_i\}, \{m_i\}, D\right) = 
\prod_i^{|I|}  \int_\mathbb{R_+} \frac{dt_i t_i^{\alpha_i-1}}{\Gamma(\alpha_i)}  \frac{\delta(1-H(t))}{\psi^{D/2}} \int_0^{+\infty} d\lambda \, \lambda^{\sum \alpha_i - D/2 \,   h_1(\Gamma) -1 } e^{-\lambda \phi/\psi}
\end{equation*}
with a small abuse of notation due to the scaling of $t_i$.
In terms of the superficial degree of divergence \(\omega\),
\begin{equation*}
\omega :=  h_1 \frac D 2 - \sum_i^{|I|} \alpha_i  ,
\end{equation*}
the integral on~\(\lambda\) gives \(\Gamma(-\omega) (\phi/\psi)^\omega\).
The presence of the hyperplane in the delta function induces a split on the domain and on the volume form 
which brings us to 
\begin{equation*}
\frac{\Gamma(-\omega)}{\prod_i \Gamma(\alpha_i)}
\int_{\mathbb{RP}^{|I|-1}_+} \Omega_H \, \,  H^{|I|} \frac{1}{\psi^{D/2}}  \Bigg(\frac{\phi}{\psi}\Bigg)^\omega\prod_i t_i^{\alpha_i-1}
\end{equation*}
written as an integral on the projective space \(\mathbb{RP}^{|I|-1}_+\) with the volume form
\(\Omega_H\) given by
\begin{equation*}
\Omega_H:= \sum_i^{|I|} (-1)^{i-1}\frac{t_i}{H} \quad d\left(\frac{t_1}{H}\right) \wedge \cdots \widehat{ d\left(\frac{t_i}{H}\right)}  \cdots \wedge d\left(\frac{t_{|I|}}{H}\right). 
\end{equation*} 
This integral is independent on the particular choice of the hyperplane thanks to its homogeneity,
since for a different choice \(H'\),  $\Omega_{H'}=(H/H')^{|I|}\Omega_H$.

In the case of the one-loop  massless diagram ($m_i=0 \quad \forall i\in I$) 
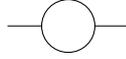
\begin{equation*}
\tikz \node[fill=white, circle, minimum size=0.7cm, draw=black]{} child[grow=east] child[grow=west];
\end{equation*}
the polynomials are very simple:
\begin{align*}
\psi = t_1 + t_2\\
\phi = t_1 t_2 p^2 .
\end{align*}
By the choice of the hyperplane $H(t)=t_2$ we get
\begin{align*}
\mathscr{I}(p^2,\alpha_1, \alpha_2, D) &:= (p^2)^{\omega} \frac{\Gamma( -\omega)}{\Gamma(\alpha_1)\Gamma(\alpha_2)}\int dt_1 dt_2 \, \delta(1-t_2)  \frac{t_1^{D/2 -\alpha_1-1}t_2^{D/2 -\alpha_2-1}}{(t_1 +t_2)^{D-\alpha_1-\alpha_2}}=\\
&=(p^2)^{ \omega} \frac{\Gamma( -\omega)}{\Gamma(\alpha_1)\Gamma(\alpha_2)} B(D/2-\alpha_1, D/2-\alpha_2)=\\
&=(p^2)^{D/2-\alpha_1-\alpha_2}\frac{\Gamma(D/2-\alpha_2)\Gamma(D/2-\alpha_1)\Gamma(-\omega)}{\Gamma(\alpha_1)\Gamma(\alpha_2)\Gamma(D/2 +\omega)}
\end{align*}
Since we are studying a model in $6$ dimensions, in the main text we will use this value.

Let us also report here the explicit calculation mentioned in section §\ref{no_vertex}. Let us consider the integral 
\begin{equation}\label{Kint_1}
\int_{\mathbb{R}^6} dq \, \left ( \frac{1}{(p+q)^2}  \right )^{1-x} 2q^\nu \left( \frac{1}{q^2} \right )^{2-y}.
\end{equation}
We can perform the Schwinger trick ~\eqref{Schwtrick} and have
\begin{equation}
\frac{2}{\Gamma(1-x)\Gamma(2-y)} \int_{\mathbb{R}_+} dt_1\int_{\mathbb{R}_+} dt_2 \, t_1^{-x}t_2^{1-y}\int_{\mathbb{R}^6} dq\, q^\nu e^{-t_1(p+q)^2-t_2q^2}.
\end{equation}
We can complete the square in the exponent
\begin{equation}
t_1(p+q)^2 +t_2q^2 = t_{12} \left (q + \frac{t_1}{t_{12}}p\right )^2 + \frac{t_1t_2}{t_{12}} p^2
\end{equation}
and then perform a change of variable $u= q+\frac{t_1}{t_{12}}p$ to get
\begin{equation}
\int_{\mathbb{R}^6} dq\, q^\nu e^{-t_1(p+q)^2-t_2q^2} = \int_{\mathbb{R}^6} du\, \left (u^\nu-\frac{t_1}{t_{12}}p^\nu\right ) e^{-t_{12}u^2} e^{-\frac{t_1t_2}{t_{12}}p^2}.
\end{equation}
In all these formula, we use the abbreviation \(t_{12} :=  t_1 + t_2 \).
The term proportional to \(u^\nu\) vanishes since it is the integral of an odd function and the
second term is a gaussian integral which gives
\begin{equation}
\int_{\mathbb{R}^6} dq\, q^\nu e^{-t_1(p+q)^2-t_2q^2} = -\pi^3p^\nu \frac{t_1}{t_{12}^4}e^{-\frac{t_1t_2}{t_{12}}p^2}.
\end{equation}
Going back to equation~\eqref{Kint_1}
\begin{equation}
\int_{\mathbb{R}^6} dq \, \left ( \frac{1}{(p+q)^2}  \right )^{1-x} 2q^\nu \left( \frac{1}{q^2} \right )^{2-y}= \frac{-2\pi^3p^\nu}{\Gamma(1-x)\Gamma(2-y)} \int_{\mathbb{R}_+} dt_1\int_{\mathbb{R}_+} dt_2 \, \frac{t_1^{1-x}t_2^{1-y} }{t_{12}^4}e^{-\frac{t_1t_2}{t_{12}}p^2}
\end{equation}
As before, we can insert the equality~\eqref{deltahyperplane} and choose the hyperplane $H(t) =
\lambda \,t_{12}$ to get
\begin{equation}
\int_{\mathbb{R}_+} dt_1\int_{\mathbb{R}_+} dt_2 \, \frac{t_1^{1-x}t_2^{1-y} }{t_{12}^4}e^{-\frac{t_1t_2}{t_{12}}p^2} =\left (p^2\right )^{x+y} \Gamma(-x-y)\int_0^1 dt_1\, t_1^{1+y} (1-t_1)^{1+x}
\end{equation}
which finally brings us to 
\begin{equation}\label{Kint}
\int_{\mathbb{R}^6} dq \, \left ( \frac{1}{(p+q)^2}  \right )^{1-x} 2q^\nu \left( \frac{1}{q^2} \right )^{2-y}= -2p^\nu \pi^3\left( p^2\right )^{x+y} \frac{\Gamma(-x-y)\Gamma(2+x)\Gamma(2+y)}{\Gamma(1-x) \Gamma(2-y)\Gamma(4+x+y)} 
\end{equation}

\section{Massless loop trick} \label{appendix2}
In this article, we are studying a massless model, so the propagators that appear in the whole text
are powers of $p^2$. With the Feynman rules for momentum space, a simple loop is the convolution of
two propagators. In the massless case though, the Fourier transform back in position space will also
be a power of \(x^2\) by homogeneity reasons and the convolution can be evaluated as a multiplication
in position space of the Fourier transforms. Apart from some \(\pi\) factors independent on~\(\alpha\), we have that
\begin{equation}
\mathcal{F}\left[\left (\frac{1}{p^2}\right)^a\right] (x) \propto \frac{\Gamma(D/2-a)}{\Gamma(a)} \left( \frac{1}{x^2}\right )^{D/2-a}
\end{equation}
so that in the case of a loop
\begin{align}
\begin{tikzpicture}
\draw (0,0) arc (0:180:.5)node [midway, above] {a};
\draw (0,0) arc (0:-180:.5)node [midway, below] {b};
\draw (0,0) -- (.2,0);
\draw (-1,0) -- (-1.2,0);
\end{tikzpicture}
\quad
&\propto
\left(\frac{1}{p^2}\right)^a \ast \left(\frac{1}{p^2}\right)^b = \mathcal{F}^{-1}  \left (
\mathcal{F}\left(\frac{1}{p^2}\right)^a \cdot  \mathcal{F}\left( \frac{1}{p^2}\right)^b \right) \\
& \propto  \frac{\Gamma(D/2-a)\Gamma(D/2-b)}{\Gamma(a)\Gamma(b)}   \mathcal{F}^{-1} \left( \frac{1}{x^2}\right)^{D-a-b} \\
&\propto \frac{\Gamma(D/2-a)\Gamma(D/2-b)\Gamma(a+b-D/2)}{\Gamma(a)\Gamma(b)\Gamma(D-a-b)} \left (\frac{1}{p^2} \right )^{a+b-D/2}
\end{align}
where we have used $\mathcal{F}(\alpha \ast \beta)= \mathcal{F}(\alpha) \cdot \mathcal{F}(\beta)$.
Repetitively using these transformations allows to evaluate the diagram appearing in
section~§\ref{fromthepropagator}.

In this section, we also have used the following construction:
starting from the diagram 
\begin{equation}
\begin{tikzpicture}
\draw (0,0) arc (0:180:.5)node [midway, above] {a};
\draw (0,0) arc (0:-180:.5)node [midway, below] {b};
\draw (0,0) -- (.2,0);
\draw (-1,0) -- (-1.2,0);
\end{tikzpicture}
\end{equation}
we can add another line of index $c$
\begin{equation}
\begin{tikzpicture} [scale=.6]
\draw (1,0) arc (0:180:1) node[midway, above] {a};
\draw (-1,0) arc (-180:0:1) node [midway, below] {c};
\draw (-1,0) -- (1,0) node [midway, above] {b};
\end{tikzpicture}
\quad
\propto 
\frac{\Gamma(D/2-a)\Gamma(D/2-b)\Gamma(D/2-c)\Gamma(a+b+c-D)}{\Gamma(a)\Gamma(b)\Gamma(c)\Gamma(3D/2 -a-b-c)} \left ( \frac{1}{p^2} \right )^{a+b+c-D}
\end{equation}
Now, if we want to consider this graph as a vacuum one, the dependence on the exterior
momentum \(p\) should disappear. We set therefore $c=D-a-b$, which is also the condition that $\omega=0$ for the
completed graph, making it logarithmically divergent. The preceding formula becomes
\begin{equation}
\frac{\Gamma(D/2-a)\Gamma(D/2-b)\Gamma(a+b-D/2)}{\Gamma(a)\Gamma(b)\Gamma(D -a-b)} \quad \frac{\Gamma(0)}{\Gamma(D/2)} 
\end{equation}
The first factor is the same we would have found if we had transformed the $a,b$ loop.  There is however a
\(\Gamma(0)/\Gamma(D/2)\) additional factor, which is infinite, reflecting the divergence of the whole
diagram.  This factor can however be interpreted as the integral
\[ \int_{\mathbb R^D} \frac{d^D u}{(u^2)^{D/2}} \]
the simplest scale invariant integral in \(D\) dimensions.  This same integral can appear either in
\(x\)-space, since the product of the three propagators gives the power \(D/2\) of \(x^2\) or in
\(p\)-space, where the combination of any two of the propagators and the last one combine to give
\((p^2)^{D/2}\). The scale invariance can be broken by fixing the momentum in any of the
propagators while giving the same number, giving another approach to the completion invariance of the
residues of propagator graphs.

In this work, we have been interested in the pole structure when one of the propagator of a completed
graph becomes scale invariant. Since the whole structure is scale invariant, the complementary of this
line becomes also scale invariant, giving two infinite factors.  We therefore understand that there should
be a pole in the evaluation of the diagram, but it is not so clear how to evaluate the residue of this
pole.  In a previous work~\cite{BeSc12}, a procedure was devised from the parametric representation, which
has the advantage of generalising to poles associated to a propagator with a power larger than \(D/2\) by
any positive integer, but another approach is possible in this simple case.  We consider the diagram in
\(x\)-space: the scale invariance is broken by fixing the distance between the two vertices while the
almost scale invariant link contributes only by its normalisation,
\(\Gamma(\varepsilon)/\Gamma(D/2-\varepsilon)\).  In the limit of vanishing~\(\varepsilon\),
\(\Gamma(\varepsilon)\) gives a pole of residue 1 while all other terms have a smooth limit.
Differentiating with respect to~\(L\) compensate the pole, so that we end up with \(1/\Gamma(D/2)\) times
the residue of the remaining scale invariant diagram.  From its scale invariance, it comes that the choice
of any two fixed vertices will give the same value for the residue, allowing for a simpler evaluation
through suitable choices of these vertices.

The same result on the residue of the pole has been previously obtained in the appendix of~\cite{GoIs85}, by a
slightly different derivation. We thank Andrei Kataev for pointing out this reference.

\section*{Potential conflict of interest}
On behalf of all authors, the corresponding author states that there is no conflict of interest.

\bibliographystyle{unsrturl}
\bibliography{../../Borel/renorm}

\begin{thebibliography}{10}

\bibitem{BrKr99}
D.~J. Broadhurst and D.~Kreimer.
\newblock Exact solutions of {D}yson--{S}chwinger equations for iterated
  one-loop integrals and propagator-coupling duality.
\newblock {\em Nucl.\ Phys.}, B 600:403--422, 2001.
\newblock \href {http://arxiv.org/abs/hep-th/0012146}
  {\path{arXiv:hep-th/0012146}}.

\bibitem{KrYe2006}
Dirk Kreimer and Karen Yeats.
\newblock {An etude in non-linear Dyson-Schwinger equations}.
\newblock {\em Nucl. Phys. Proc. Suppl.}, 160:116--121, 2006.
\newblock \href {http://arxiv.org/abs/hep-th/0605096}
  {\path{arXiv:hep-th/0605096}}, \href
  {http://dx.doi.org/10.1016/j.nuclphysbps.2006.09.036}
  {\path{doi:10.1016/j.nuclphysbps.2006.09.036}}.

\bibitem{BeSc08}
{M}arc {B}ellon and {F}idel {S}chaposnik.
\newblock Renormalization group functions for the {W}ess-{Z}umino model: up to
  200 loops through {H}opf algebras.
\newblock {\em Nucl.\ Phys.\ B}, 800:517--526, 2008.
\newblock \href {http://arxiv.org/abs/0801.0727} {\path{arXiv:0801.0727}}.

\bibitem{Be10a}
{M}arc~P. {B}ellon.
\newblock An efficient method for the solution of {S}chwinger--{D}yson
  equations for propagators.
\newblock {\em Lett.\ Math.\ Phys.}, 94:77--86, 2010.
\newblock \href {http://arxiv.org/abs/1005.0196} {\path{arXiv:1005.0196}},
  \href {http://dx.doi.org/10.1007/s11005-010-0415-3}
  {\path{doi:10.1007/s11005-010-0415-3}}.

\bibitem{BeSc12}
Marc Bellon and Fidel~A. Schaposnik.
\newblock Higher loop corrections to a {S}chwinger--{D}yson equation.
\newblock {\em Lett.\ Math.\ Phys.}, 103:881--893, 2013.
\newblock \href {http://arxiv.org/abs/1205.0022} {\path{arXiv:1205.0022}},
  \href {http://dx.doi.org/10.1007/s11005-013-0621-x}
  {\path{doi:10.1007/s11005-013-0621-x}}.

\bibitem{BeCl13}
Marc~P. Bellon and Pierre~J. Clavier.
\newblock Higher order corrections to the asymptotic perturbative solution of a
  {S}chwinger--{D}yson equation.
\newblock {\em Lett.\ Math.\ Phys.}, 104:1--22, 2014.
\newblock \href {http://arxiv.org/abs/1311.1160} {\path{arXiv:1311.1160}},
  \href {http://dx.doi.org/10.1007/s11005-014-0686-1}
  {\path{doi:10.1007/s11005-014-0686-1}}.

\bibitem{Kr2005}
Dirk Kreimer.
\newblock {Anatomy of a gauge theory}.
\newblock {\em Annals Phys.}, 321:2757--2781, 2006.
\newblock \href {http://arxiv.org/abs/hep-th/0509135}
  {\path{arXiv:hep-th/0509135}}, \href
  {http://dx.doi.org/10.1016/j.aop.2006.01.004}
  {\path{doi:10.1016/j.aop.2006.01.004}}.

\bibitem{Ye08}
Karen~Amanda Yeats.
\newblock {\em {Growth estimates for Dyson-Schwinger equations}}.
\newblock PhD thesis, Boston University, 2008.
\newblock \href {http://arxiv.org/abs/0810.2249} {\path{arXiv:0810.2249}}.

\bibitem{Wa51}
John~Clive Ward.
\newblock Renormalization theory of the inteactions of nucleons, mesons and
  photons.
\newblock {\em Phys. Rev.}, 84:897--901, 1951.
\newblock \href {http://dx.doi.org/10.1103/PhysRev.84.897}
  {\path{doi:10.1103/PhysRev.84.897}}.

\bibitem{Ward_1951}
J~C Ward.
\newblock On the renormalization of quantum electrodynamics.
\newblock {\em Proceedings of the Physical Society. Section A}, 64(1):54--56,
  jan 1951.
\newblock \href {http://dx.doi.org/10.1088/0370-1298/64/1/309}
  {\path{doi:10.1088/0370-1298/64/1/309}}.

\bibitem{Sy61}
Kurt Symanzik.
\newblock Green's functions method and renormalization of renormalizable
  quantum field theories.
\newblock In Borisov~Jak\v si\'c, editor, {\em Lectures on high energy
  physics}, pages 485--517. Gordon and Breach, 1965.
\newblock Lecture given in Herceg Novi, Montenegro, 1961.
\newblock URL: \url{https://cds.cern.ch/record/929298/files/CM-P00056764.pdf}.

\bibitem{MiYa66}
R.~L. Mills and C.~N. Yang.
\newblock Treatment of overlapping divergences in the photon self-energy
  function.
\newblock {\em Progress of Theoretical Physics Supplement}, 37-38:507--511,
  1966.
\newblock \href {http://dx.doi.org/10.1143/PTPS.37.507}
  {\path{doi:10.1143/PTPS.37.507}}.

\bibitem{BaLe77}
M.~Baker and Choonkyu Lee.
\newblock Overlapping-divergence-free skeleton expansion in non-abelian gauge
  theories.
\newblock {\em Phys. Rev. D}, 15:2201--2234, Apr 1977.
\newblock \href {http://dx.doi.org/10.1103/PhysRevD.15.2201}
  {\path{doi:10.1103/PhysRevD.15.2201}}.

\bibitem{BeCl14}
Marc~P. Bellon and Pierre~J. Clavier.
\newblock A {S}chwinger--{D}yson equation in the {B}orel plane: singularities
  of the solution.
\newblock {\em Lett.\ Math.\ Phys.}, 105:795--825, 2015.
\newblock \href {http://arxiv.org/abs/1411.7190} {\path{arXiv:1411.7190}},
  \href {http://dx.doi.org/10.1007/s11005-015-0761-2}
  {\path{doi:10.1007/s11005-015-0761-2}}.

\bibitem{BrKr2013}
Francis Brown and Dirk Kreimer.
\newblock {Angles, Scales and Parametric Renormalization}.
\newblock {\em Lett. Math. Phys.}, 103:933--1007, 2013.
\newblock \href {http://arxiv.org/abs/1112.1180} {\path{arXiv:1112.1180}},
  \href {http://dx.doi.org/10.1007/s11005-013-0625-6}
  {\path{doi:10.1007/s11005-013-0625-6}}.

\bibitem{BrKr2012}
Francis Brown and Dirk Kreimer.
\newblock {Decomposing {F}eynman rules}.
\newblock In {\em {Proceedings, 11th DESY Workshop on Elementary Particle
  Physics: Loops and Legs in Quantum Field Theory: Wernigerode, Germany, April
  15-20, 2012}}, 2012.
\newblock [PoSLL2012,049(2012)].
\newblock \href {http://arxiv.org/abs/1212.3923} {\path{arXiv:1212.3923}},
  \href {http://dx.doi.org/10.22323/1.151.0049}
  {\path{doi:10.22323/1.151.0049}}.

\bibitem{BrKr2000}
D.~J. Broadhurst and D.~Kreimer.
\newblock Towards cohomology of renormalization: bigrading the combinatorial
  {H}opf algebra of rooted trees.
\newblock {\em Comm.\ Math.\ Phys.}, 215(1):217–236, Dec 2000.
\newblock \href {http://arxiv.org/abs/hep-th/0001202}
  {\path{arXiv:hep-th/0001202}}, \href {http://dx.doi.org/10.1007/PL00005540}
  {\path{doi:10.1007/PL00005540}}.

\bibitem{Kr2003}
Dirk Kreimer.
\newblock New mathematical structures in renormalizable quantum field theories.
\newblock {\em Annals of Physics}, 303(1):179–202, Jan 2003.
\newblock \href {http://arxiv.org/abs/hep-th/0211136}
  {\path{arXiv:hep-th/0211136}}, \href
  {http://dx.doi.org/10.1016/S0003-4916(02)00023-4}
  {\path{doi:10.1016/S0003-4916(02)00023-4}}.

\bibitem{Foi2014}
Loïc Foissy.
\newblock General {D}yson–{S}chwinger equations and systems.
\newblock {\em Comm.\ Math.\ Phys.}, 327(1):151–179, Apr 2014.
\newblock \href {http://arxiv.org/abs/1112.2606} {\path{arXiv:1112.2606}},
  \href {http://dx.doi.org/10.1007/s00220-014-1941-0}
  {\path{doi:10.1007/s00220-014-1941-0}}.

\bibitem{BoPa57}
N.~Bogoliubov and O.~Parasiuk.
\newblock {\"Uber die Multiplikation der Kausalfunktionen in der Quantentheorie
  der Felder}.
\newblock {\em Acta Mathematica}, 97:227--266, 1957.
\newblock \href {http://dx.doi.org/10.1007/BF02392399}
  {\path{doi:10.1007/BF02392399}}.

\bibitem{Collins}
John~C. Collins.
\newblock {\em Renormalization: An Introduction to Renormalization, the
  Renormalization Group and the Operator-Product Expansion}.
\newblock Cambridge Monographs on Mathematical Physics. Cambridge University
  Press, 1984.
\newblock \href {http://dx.doi.org/10.1017/CBO9780511622656}
  {\path{doi:10.1017/CBO9780511622656}}.

\bibitem{KoPa2016}
Mikhail Kompaniets and Erik Panzer.
\newblock Renormalization group functions of $\phi^4$ theory in the ms-scheme
  to six loops.
\newblock {\em PoS}, LL2016:038, 2016.
\newblock \href {http://arxiv.org/abs/1606.09210} {\path{arXiv:1606.09210}},
  \href {http://dx.doi.org/10.22323/1.260.0038}
  {\path{doi:10.22323/1.260.0038}}.

\bibitem{FaChHeVe2018}
Giulio Falcioni, Konstantin Chetyrkin, Franz Herzog, and Jos Vermaseren.
\newblock {The method of global R* and its applications}.
\newblock {\em PoS}, RADCOR2017:004, 2018.
\newblock \href {http://arxiv.org/abs/1801.03024} {\path{arXiv:1801.03024}},
  \href {http://dx.doi.org/10.22323/1.290.0004}
  {\path{doi:10.22323/1.290.0004}}.

\bibitem{Sp69}
Eugene~R. Speer.
\newblock Analytic renormalization.
\newblock {\em J. Math. Phys.}, 9:1404--1410, 1969.

\bibitem{Sp71}
Eugene~R. Speer.
\newblock On the structure of analytic renormalization.
\newblock {\em Commun. Math. Phys.}, 23:23--36, 1971.

\bibitem{Speer_1976}
Eugene~R. Speer.
\newblock Dimensional and analytic renormalization.
\newblock In G.~Velo and A.~S. Wightman, editors, {\em Renormalization Theory},
  NATO Advanced Study Institutes Series, page 25–93. Springer Netherlands,
  1976.
\newblock \href {http://dx.doi.org/10.1007/978-94-010-1490-8_2}
  {\path{doi:10.1007/978-94-010-1490-8_2}}.

\bibitem{Be10}
{M}arc~P. {B}ellon.
\newblock Approximate differential equations for renormalization group
  functions in models free of vertex divergencies.
\newblock {\em Nucl.\ Phys.\ B}, 826 [{PM}]:522--531, 2010.
\newblock \href {http://arxiv.org/abs/0907.2296} {\path{arXiv:0907.2296}},
  \href {http://dx.doi.org/10.1016/j.nuclphysb.2009.11.002}
  {\path{doi:10.1016/j.nuclphysb.2009.11.002}}.

\bibitem{BeRu20a}
Marc~P. Bellon and Enrico~I. Russo.
\newblock Resurgent analysis of {W}ard--{S}chwinger--{D}yson equations.
\newblock 2020.
\newblock \href {http://arxiv.org/abs/2011.13822} {\path{arXiv:2011.13822}}.

\bibitem{Be20}
Marc~P. Bellon.
\newblock Numerators in parametric representations of {F}eynman diagrams.
\newblock in preparation, 2020.

\bibitem{Na71}
Noboru Nakanishi.
\newblock {\em Graph theory and {F}eynman integrals}, volume~11 of {\em
  Mathematics and its applications}.
\newblock Gordon and Breach, New York, 1971.

\bibitem{Gra2015}
J.~A. Gracey.
\newblock {Four loop renormalization of $\phi^3$ theory in six dimensions}.
\newblock {\em Phys. Rev.}, D92(2):025012, 2015.
\newblock \href {http://arxiv.org/abs/1506.03357} {\path{arXiv:1506.03357}},
  \href {http://dx.doi.org/10.1103/PhysRevD.92.025012}
  {\path{doi:10.1103/PhysRevD.92.025012}}.

\bibitem{PaRoSu2016}
Yi~Pang, Junchen Rong, and Ning Su.
\newblock $\phi^{3}$ theory with {F}$_{4}$ flavor symmetry in $6-2\epsilon$
  dimensions: 3-loop renormalization and conformal bootstrap.
\newblock {\em JHEP}, 12:057, 2016.
\newblock \href {http://arxiv.org/abs/1609.03007} {\path{arXiv:1609.03007}},
  \href {http://dx.doi.org/10.1007/JHEP12(2016)057}
  {\path{doi:10.1007/JHEP12(2016)057}}.

\bibitem{Pa2015}
Erik Panzer.
\newblock Algorithms for the symbolic integration of hyperlogarithms with
  applications to feynman integrals.
\newblock {\em Computer Physics Communications}, 188:148 -- 166, 2015.
\newblock \href {http://arxiv.org/abs/1403.3385} {\path{arXiv:1403.3385}},
  \href {http://dx.doi.org/10.1016/j.cpc.2014.10.019}
  {\path{doi:10.1016/j.cpc.2014.10.019}}.

\bibitem{Br86}
David~J. Broadhurst.
\newblock Exploiting the 1,440-fold symmetry of the master two-loop diagram.
\newblock {\em Z. Phys. C}, 32:249--253, 1986.

\bibitem{He66}
Klaus Hepp.
\newblock Proof of the {B}ogoliubov--{P}arasiuk theorem on renormalization.
\newblock {\em Comm.\ Math.\ Phys.}, 2:301--326, 1966.
\newblock \href {http://dx.doi.org/10.1007/BF01773358}
  {\path{doi:10.1007/BF01773358}}.

\bibitem{Zi69}
W.~Zimmermann.
\newblock Convergence of {B}ogoliubov's method of renormalization in momentum
  space.
\newblock {\em Comm.\ Math.\ Phys.}, 15:208--234, 1969.
\newblock \href {http://dx.doi.org/10.1007/BF01645676}
  {\path{doi:10.1007/BF01645676}}.

\bibitem{CoKr99}
Alain Connes and Dirk Kreimer.
\newblock {Renormalization in quantum field theory and the Riemann-Hilbert
  problem. I: The Hopf algebra structure of graphs and the main theorem}.
\newblock {\em Comm. Math. Phys.}, 210:249--273, 2000.
\newblock \href {http://arxiv.org/abs/hep-th/9912092}
  {\path{arXiv:hep-th/9912092}}, \href
  {http://dx.doi.org/10.1007/s002200050779} {\path{doi:10.1007/s002200050779}}.

\bibitem{CoKr00}
Alain Connes and Dirk Kreimer.
\newblock {Renormalization in quantum field theory and the Riemann-Hilbert
  problem. II: The beta-function, diffeomorphisms and the renormalization
  group}.
\newblock {\em Commun. Math. Phys.}, 216:215--241, 2001.
\newblock \href {http://arxiv.org/abs/hep-th/0003188}
  {\path{arXiv:hep-th/0003188}}, \href {http://dx.doi.org/10.1007/PL00005547}
  {\path{doi:10.1007/PL00005547}}.

\bibitem{EpGl73}
Henri Epstein and Vladim{\'i}r Glaser.
\newblock The role of locality in perturbation theory.
\newblock {\em Annales de l'IHP}, 19:211--295, 1973.
\newblock URL: \url{http://www.numdam.org/article/AIHPA_1973__19_3_211_0.pdf}.

\bibitem{GoIs85}
S.~G. Gorishnii and A.~P. Isaev.
\newblock An approach to the calculation of many-loop massless {F}eynman
  integrals.
\newblock {\em Theoretical and Mathematical Physics}, 62(3):232--240, 1985.
\newblock \href {http://dx.doi.org/10.1007/BF01018263}
  {\path{doi:10.1007/BF01018263}}.

\end{thebibliography}

\end{document}